# Optical Lineshape Models and the Generalized Einstein Relation between Absorption and Stimulated Emission

Aman K. Agrawal,* Jisu Ryu,*[a] and David M. Jonas[b]

Department of Chemistry, University of Colorado, Boulder, CO, 80309, USA.

*) These authors contributed equally to this work and share first authorship.
[a] Present Address: Array Labs Inc., 329 Alma St., Palo Alto, CA 94301, USA
[b] Author to whom correspondence should be addressed: david.jonas@colorado.edu

**Abstract:** Recently, Ryu *et al.* generalized Einstein's three coefficients for absorption, stimulated emission, and spontaneous emission between two quantum levels to a set of four spectra between two broadened bands. The spectra obey generalized Einstein relationships at thermal equilibrium; Einstein's relations are obtained as an approximation for line spectra. Here, the generalized Einstein relation between absorption and stimulated emission dipole-strength spectra is applied to investigate optical lineshape models. Lineshapes for the Bloch model, the stochastic model, and the semi-classical Brownian oscillator model do not obey the generalized Einstein relation and therefore fail to satisfy detailed balance with Planck blackbody radiation. The quantum Brownian oscillator model treats a harmonic quantum vibration that is bi-linearly coupled to a thermal bath of quantum harmonic oscillators which generate damping and a random force. The two-state quantum Brownian oscillator lineshape model provides lineshapes for transitions between two displaced, but otherwise identical, harmonic potential energy surfaces on which the same quantum vibration is coupled to the same thermal bath of quantum harmonic oscillators. The absorption and stimulated emission lineshapes were calculated using the quantum Brownian oscillator model in under-damped, critically damped, and over-damped cases. The thermal and reorganization energy were each varied from much less to greater than the vibrational quantum of energy. All quantum Brownian oscillator lineshapes obey the generalized Einstein relation within the numerical precision of the calculation (14 to 30 digits), suggesting this lineshape model is compatible with detailed balance. The formula giving the electric-dipole transition cross-section in terms of these lineshapes is presented.

## I. INTRODUCTION

In spectroscopy, lineshape models have been developed to simulate spectra and understand the broadening and relaxation dynamics of molecules.[1–12] Lineshape models have also been introduced that incorporate time-resolved vibrations and the interaction of these vibrations with electronic degrees of freedom and the solvent bath.[13–16] These models are applied to interpret dynamics in four-wave mixing experiments.[17–23] For a homogeneous lineshape model to be internally consistent, dynamical processes must connect transitions at different frequencies within the lineshape and the equilibrium lineshapes must satisfy detailed balance with Planck blackbody radiation.[5] Lineshapes that obey the generalized Einstein relations satisfy detailed balance with Planck blackbody radiation.[24] Here, we investigate the consistency of lineshape models with the generalized Einstein relationship between absorption and stimulated emission lineshapes.

By assuming a molecular Boltzmann distribution and equilibrium with blackbody radiation, Einstein obtained three pairwise relations among temperature-independent absorption, stimulated emission, and spontaneous emission rate constants for line spectra of a single quantum two-level system.[25] The three Einstein coefficients are specified by one *B* coefficient, one Bohr transition frequency, and one level degeneracy ratio. A generalization to broadened lineshapes that was compatible with Planck blackbody radiation remained elusive.[5] Since the Boltzmann distribution has no upper limit, lower energy system levels that are broadened by interaction with a thermal reservoir can extend above higher energy system levels. As a molecular example, highly-excited vibrational levels of the ground electronic state lie above the vibrational zero-point level of the first excited electronic state and have (in principle) non-zero probabilities in a Boltzmann distribution. If broadened levels overlap in energy, a single molecular transition direction can have both absorption and stimulated emission character so that it might also take place by spontaneous emission. A subset of system levels that equilibrates internally within the subset (but not externally with other subsets) can be treated as a broadened band. For transitions between two such broadened bands, Ryu *et al.* introduced a set of four temperature-dependent Einstein-coefficient frequency spectra for stimulated forward, spontaneous forward, stimulated reverse, and spontaneous reverse transitions. They derived a set of five pairwise single-molecule relations between these four broadband Einstein coefficient spectra by applying detailed balance between time-reversed processes under homogeneous and isotropic blackbody radiation at thermodynamic equilibrium with a single molecule.[24] This derivation assumes the system is time-reversal invariant and thus excludes rotating systems and systems in static external magnetic fields.[26] All four spectra are determined by one total *B* coefficient, one underlying lineshape that manifests differently in the four spectra, and one change in thermodynamic standard chemical potential for the transition. The theory also includes the special case of intraband transitions that start and end within the same band, which have only two Einstein coefficient frequency spectra. For an interband transition in which one band lies entirely below another to within measurement accuracy, the lower to upper spontaneous transition will practically vanish, leaving three broadened Einstein coefficient frequency spectra (absorption, stimulated emission, and spontaneous emission). The relationships between these Einstein coefficient frequency spectra differ from those for line spectra, but recover Einstein's relations as an approximation for line spectra in vacuum. In the electric-dipole approximation, the generalized Einstein relation between forward and reverse stimulated transitions also holds for dipole-strength spectra.[27] The generalized Einstein relations apply rigorously to spectra of a single molecule at thermal equilibrium.

Optical emission spectra are usually measured for an ensemble under non-equilibrium conditions, in which case the relations apply approximately to homogeneous spectra with rapid thermal quasi-equilibrium within each band. Thermal quasi-equilibrium implies that thermal equilibrium holds within each band but not necessarily between bands. For a homogeneous sample, thermal quasi-equilibrium guarantees that photoluminescence spectra will be independent of the excitation wavelength.[28] Thermal quasi-equilibrium is well known in solid-state physics, where it underpins the concept of quasi-Fermi levels.[29] If the linewidth of the absorption spectrum is significant compared to the thermal energy $k_{\mathrm{B}}T$, the generalized Einstein relation predicts that the equilibrium stimulated emission maximum will be shifted to a lower frequency than the absorption spectrum maximum. The difference between absorption and steady-state stimulated emission maxima is defined as the Stokes' shift. To obey the generalized Einstein relationship between its absorption and stimulated emission lineshapes, a lineshape model must allow the system to attain a thermal equilibrium within each band and consistently incorporate detailed balance between single-photon transitions.

The optical Bloch model for narrow lines considers homogeneous broadening from exponential relaxation, giving rise to a Lorentzian homogeneous lineshape; it may be augmented by convolution with static inhomogeneous broadening.[18,30,31] Redfield theory provides a more consistent theory of first-order relaxation.[9,32] Van Vleck and Weisskopf developed a model for pressure broadening of gas-phase microwave transitions in which the lineshape can extend to zero frequency.[4] Van Vleck and Margenau showed that the Van Vleck-Weisskopf model satisfies detailed balance between net absorption [= (absorption – stimulated emission)[33]] and spontaneous emission only for Rayleigh-Jeans blackbody radiation.[5] Van Vleck and Weisskopf claimed to reproduce both the Debye lineshape for radiofrequency dielectric relaxation in polar liquids and the optical Lorentzian lineshape as limits, but later work by Gross showed that the Van Vleck-Weisskopf model doesn't reproduce the Debye lineshape in the low-frequency limit.[34] Ben-Reuven's model reproduces the Debye lineshape as a low-frequency limit.[6] Kubo's stochastic lineshape model incorporates frequency fluctuations with time, treating homogenous and inhomogeneous broadening as dynamical in nature, separated by the timescales of the processes.[8,11] Kubo's model explains spectral diffusion and motional narrowing. None of the above models have a Stokes' shift.

Caldeira and Leggett developed the quantum Brownian oscillator model which treats a damped quantum vibration on a harmonic potential that is bi-linearly coupled to a thermal bath of quantum harmonic oscillators.[35] The quantum bath provides both damping and a random force. Grabert *et al.*[36] and Hänggi *et al*.[37] independently showed that the strongly overdamped quantum Brownian oscillator exhibits vibrational coordinate squeezing. However, it was reported for this model that the momentum variance diverges for Ohmic damping.[35–37]

The semi-classical Brownian oscillator lineshape model developed by Yan and Mukamel[13,14] and the quantum Brownian oscillator lineshape model developed by Tanimura and Mukamel[15] and by Gu, Widom and Champion,[16] describe the optical properties of a two-band system with the same Brownian oscillator in each band. The two bands have displaced harmonic vibrational potentials with the same curvature (Fig. 1). The molecular vibration has the same bi-linear coupling to a common set of bath vibrations in both bands. In these models, the harmonic bath equilibrates the molecular vibration within each band but does not couple the bands to each other. Although the harmonic and linear coupling assumptions are strong and limit their generality, they are the most reasonable analytic models available for simulating the effects of vibrational relaxation. Both models use the adiabatic approximation and the rotating-wave approximation. The semi-classical model uses a classical bath coupled to the quantum molecular vibrations, while the quantum model uses a quantum bath to describe the damping. The quantum Brownian oscillator model of ref. 16 implicitly uses the Condon approximation of a vibrational coordinate independent transition dipole, and by connecting the Condon approximation lineshape to the absorption spectrum it also implicitly assumes a single multipole order. The quantum Brownian oscillator models of refs. 15 and 16 give equivalent results for frequency-independent (Ohmic) damping in the Condon approximation. The derivation in ref. 15 invoked the fluctuation-dissipation theorem to relate the symmetrized and anti-symmetrized vibrational coordinate correlation functions. The derivation in ref. 16 explicitly invoked a detailed-balance condition [below Eq. (8) of ref. 16] for the oscillator's vibrational coordinate correlation function. The condition invoked in ref. 16 is sometimes called detailed balance (see Eq. (6.9c) in ref. 18) but it can be derived without time-reversal invariance (see ref. 38), so it is a less-restrictive condition than the detailed balance between time-reversed processes at equilibrium[26] assumed in deriving the generalized Einstein relations.[24,27] Gu et al.[16] remarked that the semi-classical Brownian oscillator model generates absorption lineshapes with significant red-edge absorption at low temperatures, suggesting a violation of detailed balance. In contrast to the semi-classical model, the quantum Brownian oscillator has negligible red-edge absorption at low temperatures.[16]

The present study goes beyond refs. 15 and 16, in discussing the stimulated emission lineshape and the connection between absorption and stimulated emission lineshapes. If a model obeys detailed balance between time-reversed single-photon transitions at equilibrium, its lineshapes should follow the generalized Einstein relation between absorption and stimulated emission. Although time-reversal invariance was not invoked in the derivations of refs. 15 and 16, the model Hamiltonian is time-reversal invariant, so the lineshapes should obey the generalized Einstein relations. We address this numerically for a broad range of parameters. We present the formula for the exact electric-dipole transition cross-section for these models.

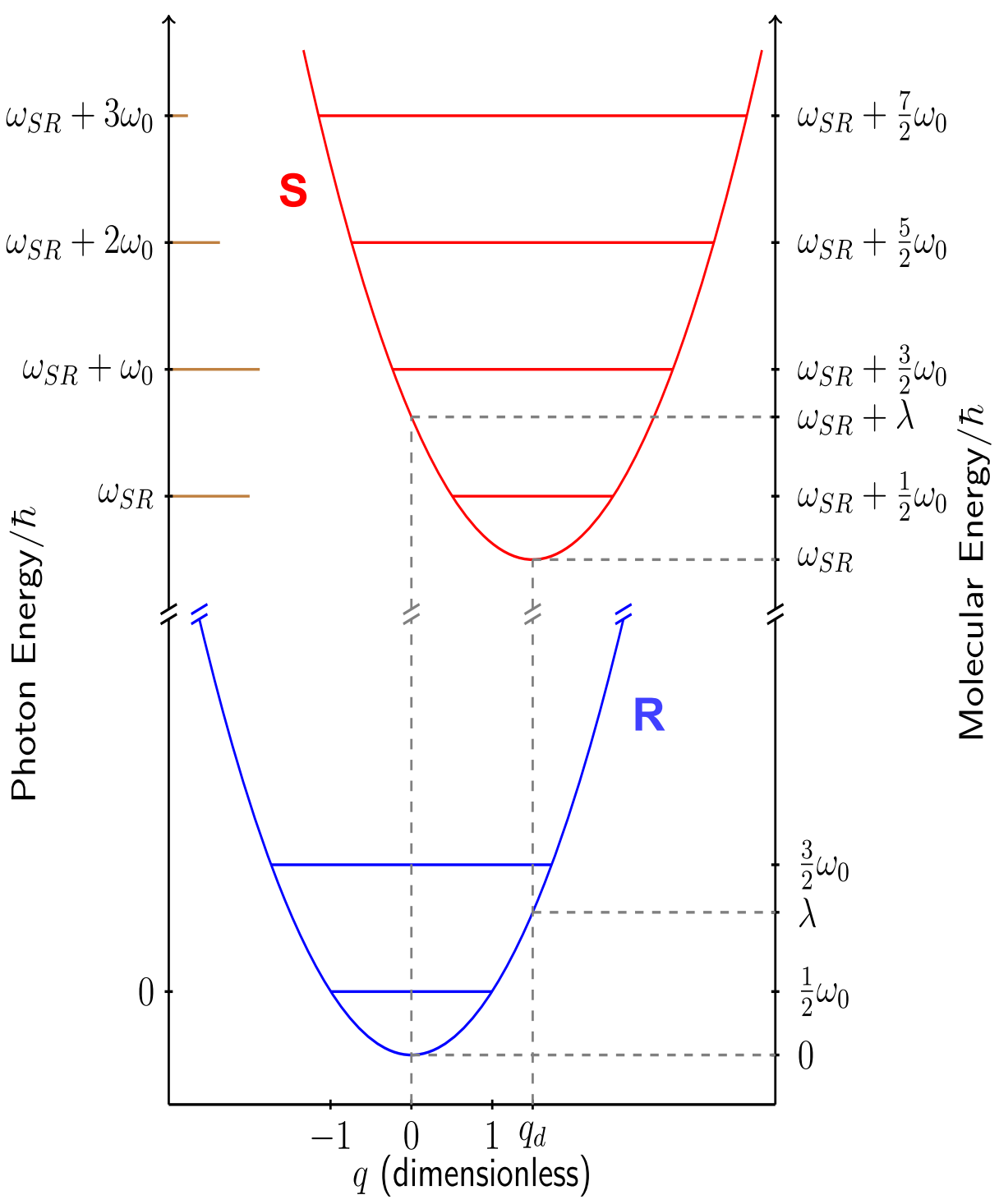

FIG. 1. Illustration of Brownian oscillator parameters and Franck-Condon factors for a transition between electronic bands *R* and *S* with quantized vibrational sublevels. The electronic potential energy curves for bands *R* and *S* have the same harmonic vibrational frequency $\omega_0$. The red upper curve for band *S* is displaced by a dimensionless normal mode displacement $q_d$ with respect to the blue lower curve for band *R*, and minima of the curves are separated by an energy gap of $\hbar\omega_{SR}$. The right vertical scale is molecular energy, while the left vertical scale is photon energy for transitions from the zero-point vibrational level on the lower potential. The orange horizontal lines on the left represent Franck-Condon factors (squares of vibrational overlap integrals) for absorption transitions from the zero-point vibrational level on the lower potential to vibrational levels on the upper potential. If the Boltzmann distribution on the lower curve populates vibrational levels that lie above the zero-point level on the upper curve, emission transitions from *R* to *S* become possible.

## II. THEORY

### A. Generalized Einstein Relations

The Einstein relations involve spectral densities that depend on the frequency units, so they are presented here with the angular frequencies most commonly used for correlation functions.[18,38,39] Einstein defined transition probabilities per unit time for three processes:[25]

absorption

$$\Gamma_{r\to s} = 2\pi B_{r\to s} u_+^{\omega}(\omega_{sr}), \qquad (1a)$$

stimulated emission

$$ {}^{B}\Gamma_{s\to r} = 2\pi B_{s\to r} u_{+}^{\omega}(\omega_{sr}), \quad (1b)$$

spontaneous emission

$$ {}^{A}\Gamma_{s\to r} = A_{s\to r}\ , \quad (1c)$$

where $\Gamma_{r\to s}$ is the conditional transition probability per unit time for a molecule in lower level $r$ to make a single-photon absorption transition to upper level $s$, $u_{+}^{\omega}(\omega)$ is the positive-frequency spectral density of electromagnetic energy per unit volume at angular frequency $\omega$, $B_{r\to s}$ is the Einstein *B* coefficient for absorption from level $r$ to level $s$, $B_{s\to r}$ is the Einstein *B* coefficient for stimulated emission from level $s$ to level $r$, and $A_{s\to r}$ is the Einstein *A* coefficient for spontaneous emission from level $s$ to level $r$. The conditional transition probability per unit time for a molecule in upper level $s$ to make a single-photon emission transition to lower level $r$ is $\Gamma_{s\to r} = {}^{A}\Gamma_{s\to r} + {}^{B}\Gamma_{s\to r}$. $\omega_{sr}$ is the absorbed and emitted frequency, which Einstein showed equals the Bohr frequency $\omega_{sr} = (E_s - E_r)/\hbar$, where $E_s$ and $E_r$ are the energies of levels $s$ and $r$, respectively, and $\hbar$ is the reduced Planck constant.

The electromagnetic energy per unit volume between positive frequencies $\omega_1$ and $\omega_2$ is given by integrating its positive-frequency spectral density:

$$U(\omega_1 \le \omega \le \omega_2) = \int_{\omega_1}^{\omega_2} u_{+}^{\omega}(\omega)\, d\omega.$$

In this paper (as in refs. 24 and 27), spectral densities[40,41] have a right superscript $\omega$ because they depend on frequency units in accord with the change of variables theorem so that $u_{+}^{\omega}(\omega = 2\pi\nu) = u_{+}^{\nu}(\nu)/2\pi$. The factor of 2π in Eqs. (1a) and (1b) arises because the spectral density for electromagnetic energy as a function of angular frequency in Eqs. (1a) and (1b) is a factor of 2π smaller than the spectral density for electromagnetic energy as a function of cyclic frequency in Einstein's corresponding rate expressions [*B*(here) = *B* (ref. 25)]. Also, the *B* coefficient is related to total dipole strength, which is the dipole-dipole correlation function at time $t = 0$, making it desirable to define *B* so that it is independent of the frequency units [in contrast to refs. 42 and 43: *B*(here) = $B^{f}$ (ref. 42) = $B^{\omega}/2\pi$ (ref. 42)].

Einstein obtained relations between these coefficients in vacuum,

$$g_r B_{r\to s} = g_s B_{s\to r}\ , \quad (2a)$$

$$A_{s\to r} = \frac{2\hbar\omega_{sr}^{3}}{\pi c^{3}} B_{s\to r}\ , \quad (2b)$$

where $g_r$ is the degeneracy of level $r$, and $c$ is the speed of light in vacuum.

For broadened spectra with possible energetic overlap between bands, the three Einstein coefficients for line spectra must be replaced by a set of four Einstein coefficient frequency spectra. None of the formulas below depend on the ambiguous question of whether *R* lies above or below *S*. Since the bands $R$ and $S$ can energetically overlap, a transition between them in one direction (e.g. forward $R \rightarrow S$) can involve both absorption and emission, and hence a molecular transition can be unclassifiable as either. However, each single-photon transition between two bands can still be classified as absorption or emission depending upon whether the transition annihilates or creates a photon. In Einstein-coefficient spectra, positive angular frequencies ($\omega > 0$) indicate photon absorption and negative angular frequencies ($\omega < 0$) indicate photon emission. For single-photon transitions between two bands $R$ and $S$ at constant pressure $p$ and temperature $T$:[24]

$$B_{R\rightarrow S}(p,T) = \frac{1}{2\pi}\int_{-\infty}^{\infty} b_{R\rightarrow S}(\omega,p,T)d\omega\,, \tag{3a}$$

$$B_{S\rightarrow R}(p,T) = \frac{1}{2\pi}\int_{-\infty}^{\infty} b_{S\rightarrow R}(\omega,p,T)d\omega\,, \tag{3b}$$

$$A_{R\rightarrow S}(p,T) = \int_{0}^{\infty} a_{R\rightarrow S}^{\omega}(-\omega,p,T)d\omega\,, \tag{3c}$$

$$A_{S\rightarrow R}(p,T) = \int_{0}^{\infty} a_{S\rightarrow R}^{\omega}(-\omega,p,T)d\omega\,, \tag{3d}$$

where $b_{R\rightarrow S}(\omega,p,T)$ and $b_{S\rightarrow R}(\omega,p,T)$ are the Einstein-coefficient frequency spectra for stimulated forward and reverse transitions between bands $R$ and $S$, $a_{R\rightarrow S}^{\omega}(-\omega,p,T)$, and $a_{S\rightarrow R}^{\omega}(-\omega,p,T)$ are the forward and reverse Einstein-coefficient spectral densities for spontaneous emission transitions between bands $R$ and $S$. For delta function line spectra between non-degenerate levels, $b_{R\rightarrow S}(\omega) = 2\pi B\delta(\omega - \omega_{sr})$ for absorption and $b_{S\rightarrow R}(\omega) = 2\pi B\delta(\omega + \omega_{sr})$ for stimulated emission. For infinitely sharp quantum energy levels, this can be viewed as taking the frequency to be the signed $(E_{\text{final}} - E_{\text{initial}})/\hbar$ instead of the positive $(E_{\text{upper}} - E_{\text{lower}})/\hbar$.

The conditional transition probabilities per unit time for a single-photon transition from band $R$ to band $S$ are

$$^{b}\Gamma_{R\rightarrow S}(u_{+}^{\omega};p,T) = \int_{0}^{\infty}[b_{R\rightarrow S}(\omega,p,T) + b_{R\rightarrow S}(-\omega,p,T)]u_{+}^{\omega}(\omega)d\omega, \tag{4a}$$

(stimulated)

$$^{a}\Gamma_{R\rightarrow S}(p,T) = \int_{0}^{\infty} a_{R\rightarrow S}^{\omega}(-\omega,p,T)d\omega\,. \tag{4b}$$

(spontaneous)

The first product inside the integral in Eq. (4a) corresponds to absorption from $R$ to $S$ and the second product corresponds to stimulated emission from $R$ to $S$. Equation (4) assumes weak molecule-field coupling, that molecules are isotropic or pseudo-isotropic through time-averaging, and that the medium is homogeneous and isotropic. The total conditional transition probability per unit time for a single-photon transition from $R$ to $S$ is the sum of Eqs. (4a) and (4b), $\Gamma_{R\to S} = {}^{a}\Gamma_{R\to S} + {}^{b}\Gamma_{R\to S}$. The corresponding equations for transition from $S$ to $R$ are obtained by interchanging the band subscripts in the above equation.

The generalized Einstein relations derived by assuming detailed balance between time-reversed processes at equilibrium with Planck blackbody radiation are

$$b_{S\to R}(-\omega,p,T) = b_{R\to S}(\omega,p,T)\exp[-(\hbar\omega - \Delta\mu^{o}_{R\to S}(p,T))/k_{\mathrm{B}}T], \quad (5a)$$

$$a^{\omega}_{S\to R}(-\omega,p,T) = \hbar\omega G^{\omega}_{+}(\omega,p,T) b_{S\to R}(-\omega,p,T), \quad (5b)$$

where $G^{\omega}_{+}(\omega,p,T)$ is the positive-frequency spectral density of electromagnetic modes per unit volume. In a weakly dispersive and (approximately) transparent medium, this spectral density is given by Eq. (7) of ref. 27 as

$$G^{\omega}_{+}(\omega,p,T) = \frac{\omega^2}{\pi^2 c^3}[n(\omega)]^2[\partial(\omega n(\omega))/\partial\omega]\theta(\omega),$$

leading to

$$a^{\omega}_{S\to R}(-\omega,p,T) = \frac{\hbar\omega^3}{\pi^2 c^3}[n(\omega)]^2[\partial(\omega n(\omega))/\partial\omega] b_{S\to R}(-\omega,p,T)\theta(\omega). \quad (5b')$$

$\Delta\mu^{o}_{R\to S}(p,T)$ is the difference in standard chemical potential per molecule between band $S$ and band $R$ [equal to the difference in standard Gibbs energy divided by Avogadro's number]. $k_{\mathrm{B}}$ is the Boltzmann constant, $n(\omega)$ is the refractive index of the medium, and $\theta(\omega)$ is the Heaviside unit step function, which is $0$ for $\omega < 0$ and $1$ for $\omega \geq 0$. Equation (5a) assumes Boltzmann statistics and ideal thermodynamic behavior of molecules (ideal gas, ideal solution, *etc.*). These relationships are valid even if the band populations are not in thermal equilibrium with each other. They only require that the initial bands are in internal thermal quasi-equilibrium with their surroundings and have the same internal temperature $T$.

The relationship between the dipole-strength spectra and the Einstein *B*-coefficient spectra in a non-magnetic, weakly dispersive, and approximately transparent medium is given by Eq. (38) of ref. 27 (in MKS units).

$$b_{R\to S}(\omega) = f^2_{local}(\omega)\left[\frac{1}{n(\omega)[\partial(\omega n(\omega))/\partial\omega]}\right]\frac{1}{6\epsilon_0\hbar^2} d^2_{R\to S}(\omega), \quad (6)$$

where $d^2$ is the dipole strength spectrum, $\epsilon_0$ is the permittivity of free space, and $f^2_{local}$ is the local electric field factor. Using Eq. (5a) and Eq. (6), we can write the relationship between the dipole-strength spectra at constant pressure and temperature as

$$d^2_{S\to R}(-\omega,p,T) = d^2_{R\to S}(\omega,p,T)\exp[-(\hbar\omega - \Delta\mu^{o}_{R\to S}(p,T))/k_{\mathrm{B}}T]. \quad (7)$$

For the Brownian oscillator model, there is no change in oscillator frequency upon electronic excitation and the coupling to the bath is unaltered, so there is no change in entropy and $\Delta\mu^{o}_{R\to S}(p,T) = \hbar\omega_{SR}$, where $\omega_{SR}$ is the 0-0 transition frequency between bands $R$ and $S$ [$= \omega^{0}_{eg}$ in Eq. (8.33c) of ref. 18].

In this paper, for functions $f(t)$ and $\hat{f}(\omega)$ related by Fourier transformation, the Fourier transform is defined as

$$f(t) = \frac{1}{2\pi}\int_{-\infty}^{\infty} \exp[-i\omega t]\, \hat{f}(\omega) d\omega\,, \qquad (8a)$$

and the inverse Fourier transform is defined as

$$\hat{f}(\omega) = \int_{-\infty}^{\infty} \exp[+i\omega t]\, f(t) dt\,. \qquad (8b)$$

A spectrum obtained by inverse Fourier transformation has $dim[\hat{f}(\omega)] = dim[f(t)] \cdot time$; in contrast to a spectral density,[41] a spectrum is independent of the frequency units, $\hat{f}(\omega = 2\pi\nu) = \hat{f}(\nu)$.

The calculations presented here used the semi-classical Brownian oscillator lineshape model of ref. 14 and the quantum Brownian oscillator lineshape model of refs. 15 and 16. In the rotating-wave approximation, the steady-state dipole-strength lineshapes for stimulated transitions from $R$ to $S$ and vice-versa, for these models are given by

$$\frac{d^2_{R\to S}(\omega)}{D^2_{R\to S}} = {}^{D}g_{R\to S}(\omega) = Re\left\{2\int_{-\infty}^{\infty} \exp[i\omega t] \exp[-i\omega_{SR}t - g(t)]\, \theta(t) dt\right\}, \qquad (9a)$$

$$\frac{d^2_{S\to R}(-\omega)}{D^2_{S\to R}} = {}^{D}g_{S\to R}(-\omega) = Re\left\{2\int_{-\infty}^{\infty} \exp[i\omega t] \exp[-i\omega_{SR}t - g^*(t)]\, \theta(t) dt\right\}, \qquad (9b)$$

where $g(t)$ is the time-domain homogeneous line-broadening function, $g^*(t)$ is the complex conjugate of $g(t)$, and $\theta(t)$ is the Heaviside unit step function which must be additionally specified as ½ for $t$ = 0 here.

$$D^2_{R\to S} \equiv \frac{1}{2\pi}\int_{-\infty}^{+\infty} d^2_{R\to S}(\omega) d\omega,$$

is the total dipole strength for the transition from *R* to *S*, and

$$\frac{1}{2\pi}\int_{-\infty}^{\infty} {}^{D}g_{R\to S}(\omega) d\omega = 1.$$

By definition, ${}^{D}g_{R\to S}$ is the area-normalized dipole-strength lineshape, and the left superscript *D* indicates a dipole-strength lineshape. [The absorption lineshape in Eq. (8.42a) and (8.B.8) of ref. 18 is $\sigma_a(\omega) = (1/2\pi)\,{}^{D}g_{R\to S}(\omega)$ and the emission lineshape in Eqs. (8.42b) and (8.B.9) is $\sigma_f(\omega) = (1/2\pi)\,{}^{D}g_{S\to R}(-\omega)$.] The normalization condition here is chosen for compatibility with the Fourier transform above, in which the $t$ = 0 point is the area of the spectrum ($D^2$ for a dipole strength spectrum, 1 for a lineshape).

## B. Semi-classical Brownian oscillator formulation

The homogeneous line-broadening function $g(t)$ for the semi-classical Brownian oscillator lineshape model is given by (Eq. (16) in ref. 17)

$$g(t) = i\lambda \int_0^t dt' M(t') + \Delta^2 \int_0^t dt' \int_0^{t'} dt" M(t") \, , \tag{10}$$

where $\lambda = (1/2)\omega_0 q_d^2$ is the reorganization energy (in angular frequency units), $\omega_0$ is the vibrational frequency, $q_d$ is the dimensionless normal mode displacement of the upper harmonic potential relative to the lower harmonic potential, $\Delta^2 = \lambda\omega_0 \coth(\beta\hbar\omega_0/2)$ is the frequency variance of the spectrum, and $\beta = 1/k_\mathrm{B}T$. This frequency variance is equal to that of the Gaussian lineshape obtained by applying the semi-classical Franck-Condon reflection principle[44] to the quantum mechanical equilibrium vibrational-coordinate probability distribution.[45] [The line-broadening functions in this paper have the same reference frequency as that in Eq. (16) of ref. 17, but a different reference frequency from Eq. (8.26) of ref. 18, so that $g_{8.26}(t) = g(t) - i\lambda t$.] In the high-temperature limit $\beta\hbar\omega_0 \ll 1$, $\Delta^2 \cong 2\lambda k_\mathrm{B}T/\hbar$. $M(t)$ is the transition-frequency correlation function,

$$M(t) = \langle \Delta\omega(0)\, \Delta\omega(t) \rangle / \langle \Delta\omega^2 \rangle, \tag{11}$$

where $\Delta\omega(t) = \omega(t) - \langle\omega\rangle$, $\langle\omega\rangle$ is the average transition frequency, and the brackets indicate an average over the ground state coordinate probability density. Because the displaced, identical-curvature harmonic oscillator model makes the transition frequency a linear function of the vibrational coordinate, the normalized transition frequency correlation functions for the damped harmonic oscillator are given by the motion of the vibrational coordinate after an initial displacement (Eq. 20 in ref. 17).

$$M(t) = \begin{cases} \exp\left(-\dfrac{\gamma t}{2}\right)\left(\cos(\omega' t) + \dfrac{\gamma}{2\omega'}\sin(\omega' t)\right) & \gamma < 2\omega_0 \, , \\ \exp\left(-\dfrac{\gamma t}{2}\right)\left(1 + \dfrac{\gamma t}{2}\right) & \gamma = 2\omega_0 \, , \\ \dfrac{s_+}{s_+ - s_-}\exp(-s_- t) - \dfrac{s_-}{s_+ - s_-}\exp(-s_+ t) & \gamma > 2\omega_0 \, , \end{cases} \tag{12}$$

where $\omega' = [\omega_0^2 - (\gamma/2)^2]^{1/2}$ and $s_\pm = \gamma/2 \pm [(\gamma/2)^2 - \omega_0^2]^{1/2}$ are real-valued and positive. The damping is characterized by $\gamma$ such that the amplitude of the under-damped oscillations decays as $\exp(-(\gamma/2)t)$.

### C. Quantum Brownian oscillator formulation

For the quantum Brownian oscillator lineshape model, the homogeneous line-broadening function $g(t)$ for a single-mode is given by (Eq. (13) in ref. 16)

$$g(t) = \frac{\lambda\omega_0^2}{2\omega'}\left(\frac{1}{\Omega_-^2}(1 - e^{-\Omega_- t}) - \frac{1}{\Omega_+^2}(1 - e^{-\Omega_+ t})\right)$$
$$+ i\frac{\lambda\omega_0^2}{2\omega'}\left\{\frac{1}{\Omega_-^2 \ \mathrm{tg}(\beta\hbar\Omega_-/2)}[1 - \Omega_- t - e^{-\Omega_- t}]\right.$$
$$\left. - \frac{1}{\Omega_+^2 \mathrm{tg}(\beta\hbar\Omega_+/2)}[1 - \Omega_+ t - e^{-\Omega_+ t}]\right\}$$
$$+ \frac{4\lambda\omega_0^2\gamma}{\hbar\beta}\sum_{n=1}^{\infty}\frac{1 - \omega_n t - e^{-\omega_n t}}{\omega_n[(\omega_0^2 + \omega_n^2)^2 - \gamma^2\omega_n^2]}.$$

(13)

For complex-valued $z$, $\mathrm{tg}(z) = (e^{iz} - e^{-iz})/i(e^{iz} + e^{-iz})$. In Eq. (13) here, the damping constant $\gamma$ is defined in the same way as below Eq. (12), which is different from in ref. 16: here $\gamma/2 = \gamma$ (ref. 16). For an underdamped oscillator ($\gamma < 2\omega_0$), $\Omega_\pm = \gamma/2 \pm i\omega'$ with $\omega'$ real-valued as in Eq. (12); for an overdamped oscillator ($\gamma > 2\omega_0$), $\Omega_\pm = \gamma/2 \mp [(\gamma/2)^2 - \omega_0^2]^{1/2}$ so that $\Omega_\mp = s_\pm$ in Eq. (12), and $\omega' = i[(\gamma/2)^2 - \omega_0^2]^{1/2}$ is imaginary with a positive imaginary part. In Eq. (13), the 1st term in large parentheses is purely imaginary, the 2nd term in large braces is purely real, and the 3rd term that contains a sum over the Matsubara frequencies $\omega_n = 2\pi n/\beta\hbar$ is purely real. The imaginary part of $g(t)$ is the same for the classical bath model and the quantum bath model, but the real part of $g(t)$ differs.

**D. Transition cross-sections**

To allow the possibility that a single molecular transition direction may not be purely absorptive or emissive, refs. 24 and 27 derived Beer's Law in the following form

$$\frac{I^\omega(z;\omega)}{I^\omega(z = 0;\omega)} = \exp\left[-\sum_{R,S} N_R\sigma_{R\to S}(\omega,p,T)z\right], \tag{14}$$

where $I^\omega$ is the spectral irradiance, *z* is the distance inside the sample, $N_R$ is the molecular number density in the initial band $R$, and $\sigma_{R\to S}$ is a *signed* transition cross-section for the $R$ to $S$ transition. The signed *R* to *S* transition cross-section used here is positive for frequencies at which absorption dominates the *R* to *S* transition and negative for frequencies at which stimulated emission dominates the *R* to *S* transition, so that the above Beer's law result does not depend on whether *R* lies below or above *S*. The sum is over all initial bands $R$ and final bands $S$ and includes intraband transitions with $R = S$. (If the bands *R* and *S* are molecular electronic states with *R* below *S*, the intraband transition *R* to *R* contains the vibrational spectrum of band *R*, and its cross section gives the *net* vibrational absorption.) For a non-magnetic medium, the transition cross-section for the $R$ to $S$ transition is related to its dipole-strength spectrum by (MKS units)[27]

$$\sigma_{R\to S}(\omega,p,T) = \frac{\hbar\omega}{n(\omega)c}\frac{1}{6\epsilon_0\hbar^2}f_{local}^2(\omega)[d_{R\to S}^2(\omega,p,T) - d_{R\to S}^2(-\omega,p,T)]. \tag{15}$$

For positive frequency $\omega$, $d^2(\omega)$ represents absorption and $d^2(-\omega)$ represents stimulated emission. The expression for the transition cross-section for the $S$ to $R$ transition is obtained by interchanging $R$ and $S$. Transition cross-sections are positive for net absorption and negative for net stimulated emission. This formulation in Eq. (14) and (15) is different from the usual formula in which all cross-sections are defined to be positive.[30] In the usual Beer's law formula with positive cross-sections, one has to insert positive and negative signs in front of absorption and stimulated emission cross-sections, respectively. Here, Eq. (14) and Eq. (15) algebraically distinguish net absorption from net stimulated emission through the sign of the transition cross-section itself. This simply allows for the possibility that a single molecular transition direction (for example, *R* to *S*) can be dominated by absorption for some frequencies and stimulated emission for other frequencies.

When considered over the entire real frequency axis, each transition cross-section is an even function of frequency. This even symmetry is required by the real-valued nature of time-domain optical electric fields. For interband transitions, the two terms in Eq. (15) are equivalent to the positive and negative frequency terms in the exact rotating-wave decomposition of the impulse response and susceptibility.[24,46] This decomposition becomes exact for vanishing bath-induced relaxation between distinct initial and final bands, so the subtraction in Eq. (15) repairs the rotating-wave approximation in the Brownian oscillator models.

The treatment of transition cross-sections here is significantly different from that in ref. 18, where the "spectral density $J(\omega)$" (correlation function terminology)[41] involves absorptive transitions (from *R* to *S* here) at positive frequency and emissive transitions (from *S* to *R* here) at negative frequency [see Eq. (6.8a) and Fig. 6.1b of ref. 18]. That combined $J(\omega)$ depends on the equilibrium populations of two different bands. Here, the transition cross-section for each molecular transition direction [*e.g.* from *R* to *S* in Eq. (15)] is treated separately, and each transition cross-section is multiplied by the population of the initial band in Beer's law. At equilibrium, the two approaches agree and their difference vanishes through a regrouping of terms and use of an algebraic sign (instead of a manual sign change) for stimulated emission. One advantage of the different approach here is that, after each populated initial band reaches internal quasi-equilibrium, it allows non-equilibrium populations for different bands in Beer's law. This ability to treat non-equilibrium band populations is valuable even if all transitions can be classified as absorptive or emissive.

## III. CALCULATIONS

The high accuracy of the calculations presented here is used to test whether lineshape models obey the generalized Einstein relations, which involve exponential amplification of low amplitude parts of the lineshape. Accurate numerical calculations can prove that a lineshape model does not satisfy the generalized Einstein relations, but can only suggest that they might be satisfied. The code for the calculation of optical lineshapes was adapted from the routines available in the Electronic Physics

Auxiliary Publication for refs. 47 and 48 by rewriting for quadruple precision calculations in Fortran 95/2003. The code was compiled using the Intel(R) Visual Fortran Compiler 19.1.1.216 [Intel(R) 64]. The program was run under a 64-bit Windows 10 Professional operating system on a computer with a 1.80GHz Core i7 processor and 16 GB of random access memory (RAM). The calculations were carried out on a grid where each grid point is a complex quadruple-precision number. The absorption and emission spectra were calculated by performing a complex-valued, quadruple-precision, inverse Fourier transform using the FFT subroutines four1 and fourrow defined in ref. 49. The time grid for the FFT starts at $t$ = 0 and all times on the grid are treated as positive. At the largest positive time for the calculations here, the largest value of $|\exp[-g(t)]|$ is $3.7 \times 10^{-11}$. To prevent baseline offset in the frequency domain spectrum, the first data point at $t$ = 0 in the time domain is multiplied by one-half (as prescribed by the form of the Heaviside unit step function below Eq. 9) before the discrete Fourier transform.[50,51] This approach is supported by the analytical result for a point of discontinuity, where the function retrieved by Fourier transformation and inverse Fourier transformation is equal to the average of the limits of the discontinuous function as approached from the two sides (see section 1.9 of ref. 52). The fundamental constants used in the code are from ref. 53 and $\pi$ was taken to 31 digits from ref. 54. In calculating the line-broadening function for a critically damped oscillator using the quantum Brownian oscillator model, the 1$^{st}$ term in large parentheses and the 2$^{nd}$ term in large braces in Eq. (13) give $0/0$. Hence, these terms in $g(t)$ were re-formulated using L'Hopital's rule. The subroutines used in calculating $g(t)$ for the underdamped, critically damped, and over-damped quantum Brownian oscillators were *exactly* those available in the Electronic Physics Auxiliary Publication for ref. 47. The functions for the Matsubara sum were rewritten with more complete convergence criteria and are provided here in Supplementary Material.

**Convergence Test:** Three types of convergence tests were performed to estimate the error in the calculation of optical lineshapes arising from using a finite number of terms in the Matsubara sum, a finite time-range grid,[55] and a non-zero time-step.[55]

First, in calculating optical lineshapes for the quantum Brownian oscillator model, the sum over the Matsubara frequencies [3$^{rd}$ term in Eq. (13)] always extends beyond the least Matsubara frequency with $\omega_n > \gamma$ so that it extends beyond the change in sign of the denominator. The numerator can be split into three sums, with lineshape convergence most sensitive to the sum from the second ($\omega_n t$) term; factoring out the time $t$, this sum was converged by evaluating it until the magnitude of the difference between the sum of $N$+100 and $N$ Matsubara terms is less than a tolerance set at $10^{-28}$ for the calculations presented here to allow for multiplication by large $t$. The sum from the first (constant) term in the numerator was also evaluated only once, with a tolerance set at $10^{-25}$. The sum from the third (exponential) term in the numerator must be evaluated for each $t$ but converges rapidly with a tolerance set at $10^{-25}$. These convergence criteria were chosen to avoid negative regions in the absorption lineshape, converge beyond the error dictated by the time resolution of the discrete

FFT, and, most critically, minimize error from exponential amplification when testing the generalized Einstein relation.

Second, the effect of the finite time range was evaluated by comparing absorption lineshapes calculated using two different grid sizes. The larger grid has double the time range of the smaller grid, but both have the same time step. Since the time-step size is the same, both spectra have the same frequency range, but the larger grid has twice the frequency resolution because its time range is doubled. The relative error between spectra was determined by subtracting the spectra at common frequencies and then dividing by the maximum of the spectrum.

Third, the effect of using discrete time points was determined by comparing two absorption spectra calculated on grids with time-step sizes differing by a factor of 2 but the same time range. Since the time range is the same, the frequency resolution is the same, but the spectrum on the larger grid has twice the frequency range because it has half the time-step size. The two spectra were compared by first cropping the spectrum calculated on the larger grid to the same frequency range as the spectrum calculated on the smaller grid. The relative difference spectrum was then determined by subtracting the two spectra and dividing by the global maximum of the spectrum. The maximum of the absolute value of the relative difference spectrum is reported as the global precision of the calculation. The accuracy of these calculations is always limited by FFT aliasing[56–58] wrap-around errors on the low-frequency side of absorption lineshapes and the high-frequency side of stimulated emission lineshapes. On a logarithmic scale, Fig. 2(c) shows that the high-frequency side of these model absorption lineshapes decays extremely slowly and is thus wrapped around to frequencies below the peak by FFT aliasing (see Fig. S1) – the effect in stimulated emission is a mirror image. These wrap-around errors generate a floor that limits numerical tests of the generalized Einstein relation. Figure 2(d) shows the unwrapped lineshape, but it should be kept in mind that the high-frequency wing of the lineshape is wrapped around an infinite number of times, contaminating the lineshape everywhere; the converged result in Fig. 2(c) shows that this wrap-around error is largest at the low-frequency side of the absorption lineshape.

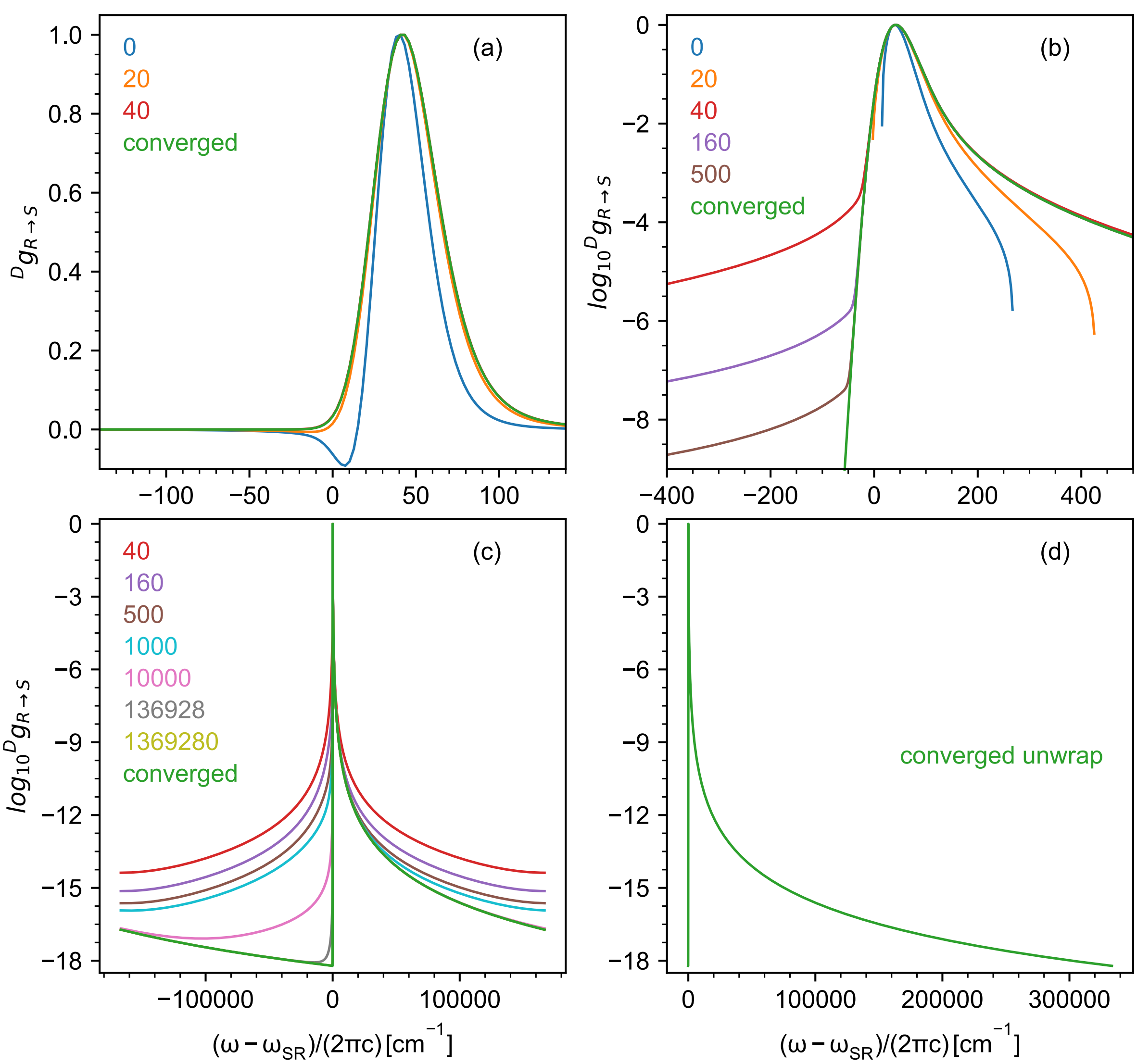

FIG. 2. Effect of the number of terms included in the Matsubara sum in Eq. (13) on the absorption lineshape for an over-damped quantum Brownian oscillator with vibrational frequency $\omega_0/(2\pi c) = 50\ \mathrm{cm}^{-1}$, damping constant $\gamma/(2\pi c) = 600\ \mathrm{cm}^{-1}$ and reorganization energy $\lambda/(2\pi c) = 50\ \mathrm{cm}^{-1}$ at a temperature of 4K. A grid of 131072-time steps with 0.1 fs spacing yielded global precision of $10^{-17}$ for the converged absorption lineshape. The converged lineshape required 13692800 terms in the Matsubara sum. (a) plots the height-normalized absorption lineshapes. The lineshape for 40 terms is hidden under the converged lineshape at this scale. Height-normalized absorption lineshapes are shown on a logarithmic scale over a larger frequency range in (b) and the full frequency range in (c). The absence of data points in (b) for 0 and 20 terms indicates that the calculated absorption lineshape is negative at those frequencies. At this scale, the lineshape for 1369280 terms is hidden under the converged lineshape in (c). The converged lineshape in (c) reveals numerical FFT aliasing – the lineshape to the left of the minimum is a continuation of the high-frequency wing that has been wrapped around to low frequency. (d) shows the unwrapped converged absorption lineshape.

Gu *et al.*[16] noted that the Matsubara sum provides a major quantum correction that is critical for generating asymmetric lineshapes at low temperatures where hot band transitions vanish (Fig. S2 in the SI). Figure 2 shows an additional effect of the Matsubara sum on the lineshape for a strongly over-damped quantum oscillator ($\omega_0/(2\pi c) = 50\ \mathrm{cm}^{-1}$, $\gamma/(2\pi c) = 600\ \mathrm{cm}^{-1}$, $\lambda/(2\pi c) = 50\ \mathrm{cm}^{-1}$) at a temperature of 4K. At this temperature the thermal energy is much less than a quantum of vibrational energy and strong over-damping was chosen to reveal how lineshapes depend on strong interactions with a quantum bath obeying Bose statistics. When too few terms are included in the Matsubara sum, the optical lineshape has numerically significant negative features on both the low- and high-frequency sides [not visible on the linear scale for 20 terms in Fig. 2(a), the curves disappear for negative absorption in Fig. 2(b)]. For this model, more than 20 terms in the Matsubara sum were needed to avoid negative features in the optical lineshape. Because the generalized Einstein relation prediction of stimulated emission from absorption exponentially amplifies the low-frequency region of the absorption lineshape, any numerical test of this prediction will be limited by lineshape accuracy there. Figure 2(c) illustrates how strongly the number of terms in the Matsubara sum affects the low-frequency frequency region of the absorption lineshape. The maximum absolute difference between the converged lineshape and the lineshape with 1369280 terms divided by the global maximum of the spectrum is about $10^{-16}$. We needed 13692800 terms [corresponding to a maximum Matsubara frequency of 45054 rad/fs] for the above calculation to converge within the numerical global precision of $10^{-17}$ for the calculation. This converged result is decreasing down to ~137 $\mathrm{cm}^{-1}$ below $\omega_{SR}/(2\pi c)$, where FFT wraparound error limits the numerical accuracy.

**Comparison to Analytic Theory:** Convergence tests above quantify precision of spectral lineshape calculations. A comparison of the numerically calculated optical lineshape using the discrete Fast Fourier transform (FFT) method to analytic theory can be used to quantify absolute accuracy. For the homogeneous Gaussian lineshape formula, the line-broadening function is $g(t) = i\lambda t + (1/2)\Delta^2 t^2$ where $\Delta^2 = 2\lambda k_B T/\hbar$ is the variance of the Gaussian lineshape. Eq. (9) is used to calculate optical lineshapes. Figure S3 of the SI shows the lineshape calculated from the Gaussian line-broadening function with $\lambda/(2\pi c) = 50\ \mathrm{cm}^{-1}$ at 100K. Comparison between the analytical Gaussian lineshape and the numerical lineshape calculated from the Gaussian line-broadening function using the FFT method demonstrates a global accuracy of better than $10^{-31}$.

Homogeneous Gaussian lineshapes with variance $\Delta^2$ and Stokes' shift $(2\lambda)$ related by $\Delta^2 = (2\lambda)k_\mathrm{B}T/\hbar$ obey the generalized Einstein relations exactly.[59] Although Gaussian lineshapes can be approximately obtained in various high-temperature limits of the strongly overdamped Brownian oscillator lineshape model, the numerical tests here simply start from the analytical Fourier transforms of the Gaussian lineshapes in the time domain and rely only on the above exact analytical result for Gaussian lineshapes to test calculation accuracy. We used the homogeneous Gaussian lineshape formula to test the accuracy of FFT calculations for prediction of the stimulated emission lineshape from the absorption lineshape using the generalized Einstein relation, Eq.

(7). Figure S3(b) shows the difference between the stimulated emission lineshape calculated from the Gaussian line-broadening function and the 'GER-predicted' stimulated emission lineshape determined by applying the generalized Einstein relation (GER) to the absorption lineshape calculated from the Gaussian line-broadening function. For height-normalized lineshapes, the maximum generalized Einstein relation difference in Fig. S3(b) is less than $10^{-27}$ at a frequency where the height-normalized absorption lineshape is $10^{-35}$.

The relative error in the GER-predicted stimulated emission lineshape was determined by subtracting it from the calculated stimulated emission lineshape and dividing the difference by the calculated stimulated emission lineshape at each frequency. At each frequency, the relative error in the GER-predicted stimulated emission lineshape is approximately equal to the relative error in the absorption lineshape. At temperatures of 100 K and above, the absorption lineshapes are still decreasing at -750 $cm^{-1}$ and can be tested for convergence there. For example, for the spectrum in Fig. S3, the height-normalized absorption lineshape is ~$10^{-20}$ there, the relative error in the absorption lineshape is ~$10^{-14}$ there, and the generalized Einstein relation prediction is accurate to within this relative error.

The calculated Brownian oscillator lineshapes were also checked for the parameters in Fig. 1 of ref. 16 by visually overlapping them [Fig. S4 reproduces Fig. 1a of ref. 16, and Fig. S5 and S6 reproduce Fig. 1b of ref. 16]. As a check on unexpected results, the semi-classical Brownian oscillator lineshapes calculated using the complex-valued line-broadening function in Eq. (14) of ref. 16 reproduced the lineshapes calculated using separate real and imaginary line-broadening functions from Eq. (10) [FORTRAN functions realg and imagg from the Appendix to ref. 21] to within $10^{-15}$.

## IV. RESULTS

The generalized Einstein relation was first tested on the semi-classical Brownian oscillator model of Yan and Mukamel.[14] Figure 3 shows the semi-classical Brownian oscillator lineshape for an over-damped oscillator with $\omega_0/(2\pi c) = 50\ \mathrm{cm}^{-1}$, $\gamma/(2\pi c) = 600\ \mathrm{cm}^{-1}$, and $\lambda/(2\pi c) = 50\ \mathrm{cm}^{-1}$ at 100K. The oscillator parameters are the same as those used in Fig. 2 here and in Fig. 1 of ref. 16, but the temperature is between the two shown in ref. 16. Figure 3(a) shows that the GER-predicted stimulated emission lineshape does not agree with the stimulated emission lineshape calculated from the model. The disagreement is 3% at the emission peak, which is significantly larger than the $10^{-16}$ global precision and experimentally detectable.[60] This demonstrates that the semi-classical Brownian oscillator model[14] does not obey the generalized Einstein relation. Since it is a semi-classical model, it is expected to agree better with the generalized Einstein relation at a higher temperature. The lineshape calculation results at 300K ($k_\mathrm{B}T/(\hbar\omega_0) \approx 4$) with the same Brownian oscillator parameters agree with this expectation (Fig. S5 in the SI). One striking feature in Fig. 3(a) is the negative GER-predicted stimulated emission lineshape below the 0-0 transition frequency. This implies that the model's absorption lineshape is incorrectly negative at those frequencies, a feature seen in both Fig. 3(b) and in Fig. S5(b). The semi-classical

Brownian oscillator model produces absorption lineshapes that have numerically significant low-amplitude negative features. The negative feature for this strongly over-damped semi-classical Brownian oscillator is reminiscent of the negative feature in Fig. 2 for the strongly over-damped quantum Brownian oscillator with too few terms in the Matsubara sum. Figure S2 does not show negative features when the Matsubara sum is neglected for a less strongly over-damped quantum Brownian oscillator. We found these negative features from the semi-classical Brownian oscillator model surprising because the quantum Golden Rule gives non-negative dipole-strength spectra.[27]

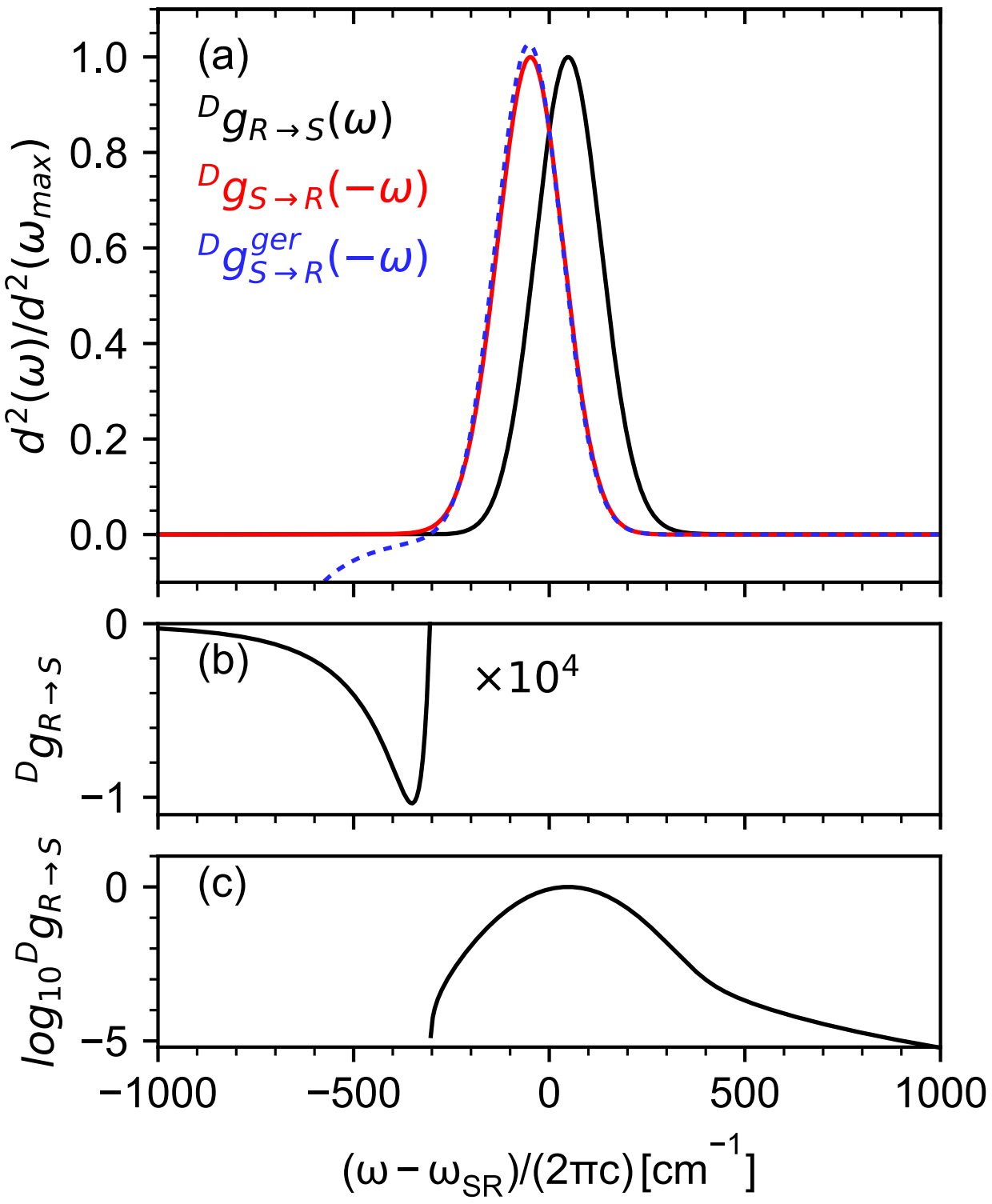

FIG. 3. Test of the semi-classical Brownian oscillator model against the generalized Einstein relation for an over-damped oscillator with vibrational frequency $\omega_0/(2\pi c) = 50\ \mathrm{cm}^{-1}$, damping constant $\gamma/(2\pi c) = 600\ \mathrm{cm}^{-1}$ and reorganization energy $\lambda/(2\pi c) = 50\ \mathrm{cm}^{-1}$ at a temperature of 100K. A grid of 131072-time steps with 0.1 fs spacing yielded a global precision of $10^{-16}$. (a) shows lineshapes for absorption (black) and stimulated emission (red) along with the GER-predicted stimulated emission lineshape (dashed blue). (b) shows the negative part of the absorption lineshape, which is not directly visible in (a), on a rescaled vertical axis. (c) shows the positive part of the absorption lineshape on a logarithmic scale.

Next, the generalized Einstein relation was used to test the quantum Brownian oscillator model of refs. 15 and 16. Figure 4 shows lineshapes calculated using the quantum Brownian oscillator with the same parameters as Fig. 3. The GER-predicted stimulated emission lineshape visually agrees with the model's stimulated emission lineshape. The relative error in the GER-predicted stimulated emission lineshape [calculated by dividing the difference shown in Fig. 4(b) by the stimulated emission shown in Fig. 4(c)] is approximately equal to the relative error in the absorption lineshape [calculated by dividing the precision of absorption (not shown) by the absorption shown in Fig. 4(c)]. This observation agrees with the observation of relative errors in GER-prediction and absorption lineshape using the Gaussian lineshape. So, in this calculation the quantum Brownian oscillator lineshape model numerically agrees with the generalized Einstein relation. The error in the GER-predicted stimulated emission lineshape increases as frequency decreases. This arises from exponential amplification of the wrap-around error in the absorption lineshape generated from multiplication by an exponentially increasing function as the frequency decreases. The absorption lineshape is positive across the whole frequency range (as required), unlike the semi-classical Brownian oscillator lineshape. Figure S6 shows lineshapes for the same over-damped quantum oscillator at 300K, where the thermal energy is greater than a quantum of vibrational energy. The GER-predicted stimulated emission lineshape agrees with the model's stimulated emission lineshape at 300K for the above set of oscillator parameters.

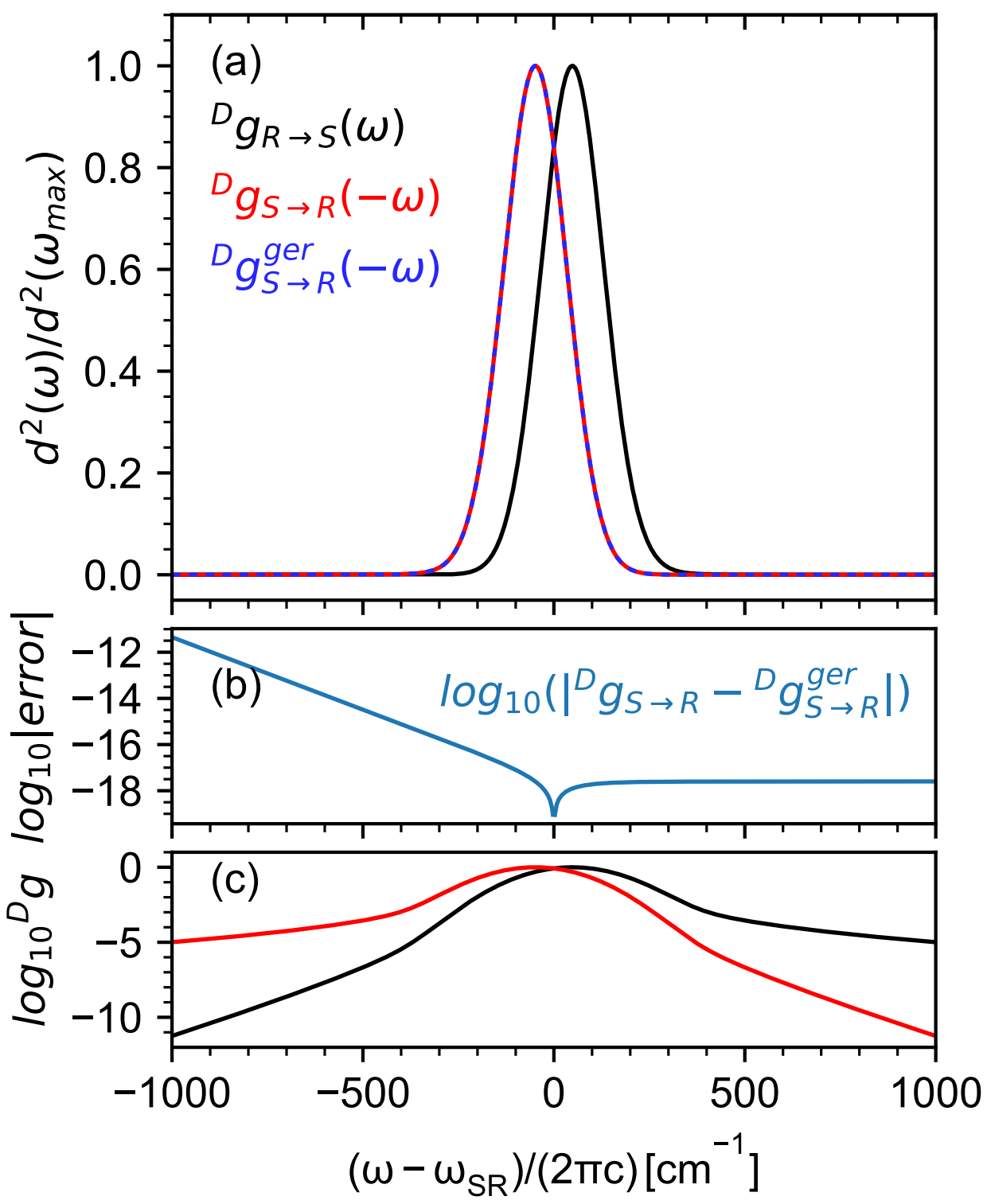

FIG. 4. Test of the quantum Brownian oscillator model against the generalized Einstein relation for an over-damped quantum Brownian oscillator with the same parameters as Fig. 3 [vibrational frequency $\omega_0/(2\pi c) = 50\ \mathrm{cm}^{-1}$, damping constant $\gamma/(2\pi c) = 600\ \mathrm{cm}^{-1}$ and reorganization energy $\lambda/(2\pi c) = 50\ \mathrm{cm}^{-1}$] at the same temperature of 100K. A grid of 131072-time steps with 0.1 fs spacing yielded global precision of $10^{-16}$. (a) shows lineshapes for absorption (black) and stimulated emission (red) along with the GER-predicted stimulated emission lineshape (dashed blue). (b) shows the difference between the stimulated emission lineshape and the GER-predicted stimulated emission lineshape on a logarithmic scale. (c) shows absorption and stimulated emission lineshapes on a logarithmic scale.

The numerical agreement between the quantum Brownian oscillator model and the generalized Einstein relation was tested by calculating lineshapes for a range of oscillator parameters and temperatures. Figure 5 shows lineshapes for the over-damped quantum oscillator of Fig. 4 but at 2K. These parameters are the same as those used in Fig. 1(a) of ref. 16 and visually reproduce Fig. 1(a) of ref. 16. At this temperature the thermal energy is much less than a quantum of vibrational energy. The absorption lineshape at low temperature is asymmetric, much sharper than the absorption lineshape at 100K, and demonstrates the absence of hot-band transitions at 2K discussed previously by Gu *et al.*[16]. The GER-predicted stimulated emission lineshape quantitatively agrees with the model's stimulated emission lineshape even at 2K temperature. The generalized Einstein relation holds even though the overlap between absorption and stimulated emission lineshapes is practically negligible. The divergence in the GER-predicted stimulated emission lineshape below about $-60\ \mathrm{cm}^{-1}$ arises because the absorption is smaller (Fig. 5(b)) than the $10^{-17}$ global precision (limited by FFT wrap-around error), and the absorption is multiplied by an exponential function that increases with decrease in frequency.

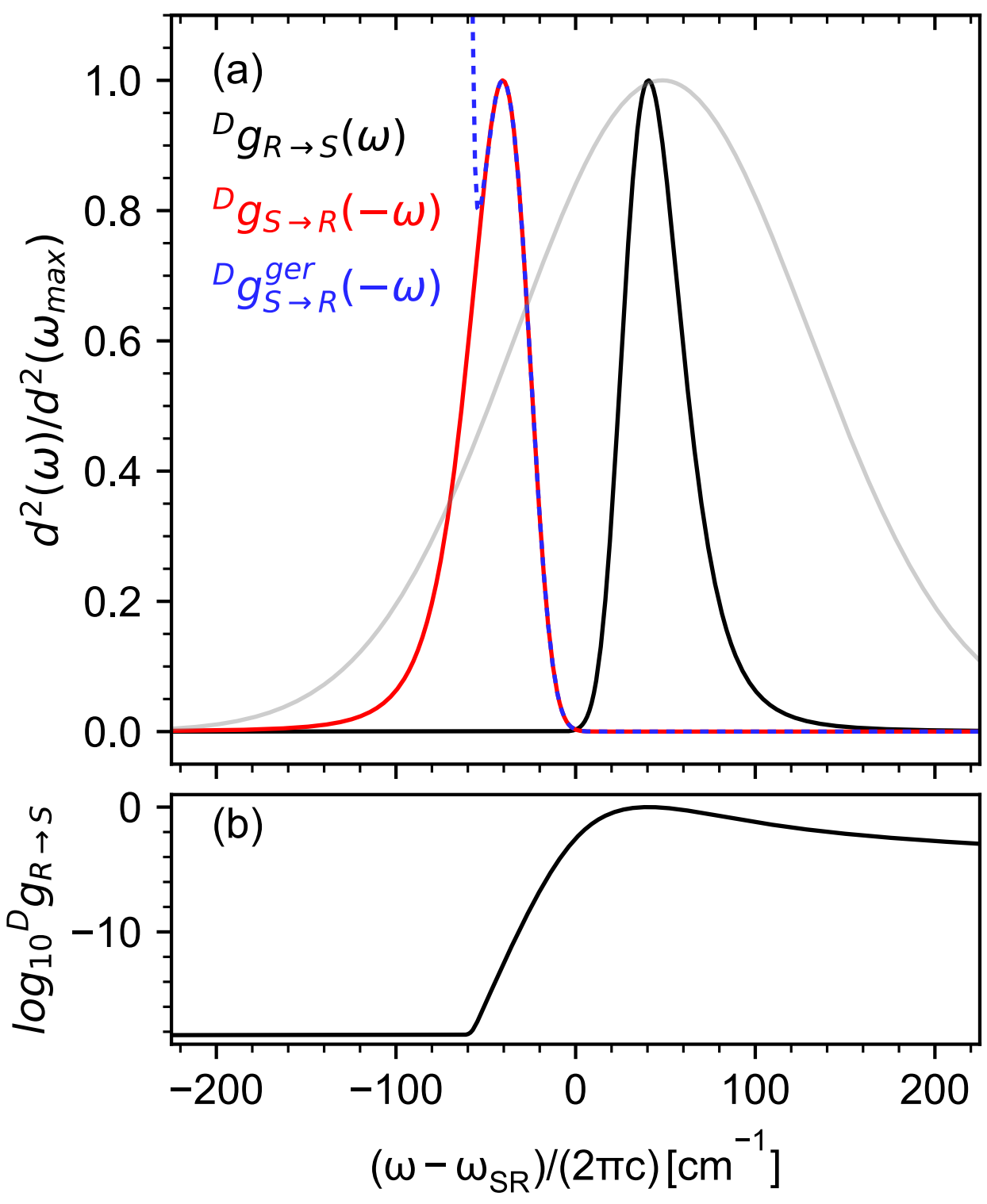

FIG. 5. Test of the quantum Brownian oscillator model against the generalized Einstein relation for the same over-damped oscillator parameters as Figs. 2-4 [vibrational frequency $\omega_0/(2\pi c) = 50\ \mathrm{cm}^{-1}$, damping constant $\gamma/(2\pi c) = 600\ \mathrm{cm}^{-1}$ and reorganization energy $\lambda/(2\pi c) = 50\ \mathrm{cm}^{-1}$] but at a temperature of 2K. A grid of 262144-time steps with 0.1 fs spacing yielded a global precision of $10^{-17}$. Panel (a) shows lineshapes for absorption (black) and stimulated emission (red) along with the GER-predicted stimulated emission lineshape (dashed blue). The grey curve is the absorption lineshape at 100K from Fig. 4. Panel (b) shows the absorption lineshape on a logarithmic scale.

Figure 6 shows lineshapes for a critically damped quantum Brownian oscillator at 2K with parameters ($\omega_0/(2\pi c) = 70\ \mathrm{cm}^{-1},\ \lambda/(2\pi c) = 30\ \mathrm{cm}^{-1}$) corresponding to the correlated vibration parameters used in ref. 61 for the FMO dimer model that roughly approximates two excitons in the Fenna-Matthews-Olson (FMO) light-harvesting pigment-protein complex. For these parameters, the reorganization energy is less than a quantum of vibrational energy and the thermal energy is much less than a quantum of vibrational energy. The Stokes' shift, calculated as the frequency difference between the absorption and stimulated emission maxima, is $\sim 2.5\ \mathrm{cm}^{-1}$ which is much less than the reorganization energy. Again, the GER-predicted stimulated emission lineshape quantitatively agrees with the model's stimulated emission lineshape. The divergence below about $-60\ \mathrm{cm}^{-1}$ in the GER-predicted stimulated emission lineshape arises for the same wrap-around/exponential amplification reasons discussed for Fig. 5.

The undamped limit provides perspective on this Stokes' shift. In the undamped limit, the 0-0 transition would be the strongest in both absorption and emission for this reorganization energy, yielding zero Stokes' shift. The vibrational structure of the undamped oscillator's electronic transitions is completely washed out for this critically damped oscillator, yet the Stokes' shift remains more than an order of magnitude smaller than twice the reorganization energy. This low temperature result is compatible with the generalized Einstein relation between equilibrium absorption and emission spectra.

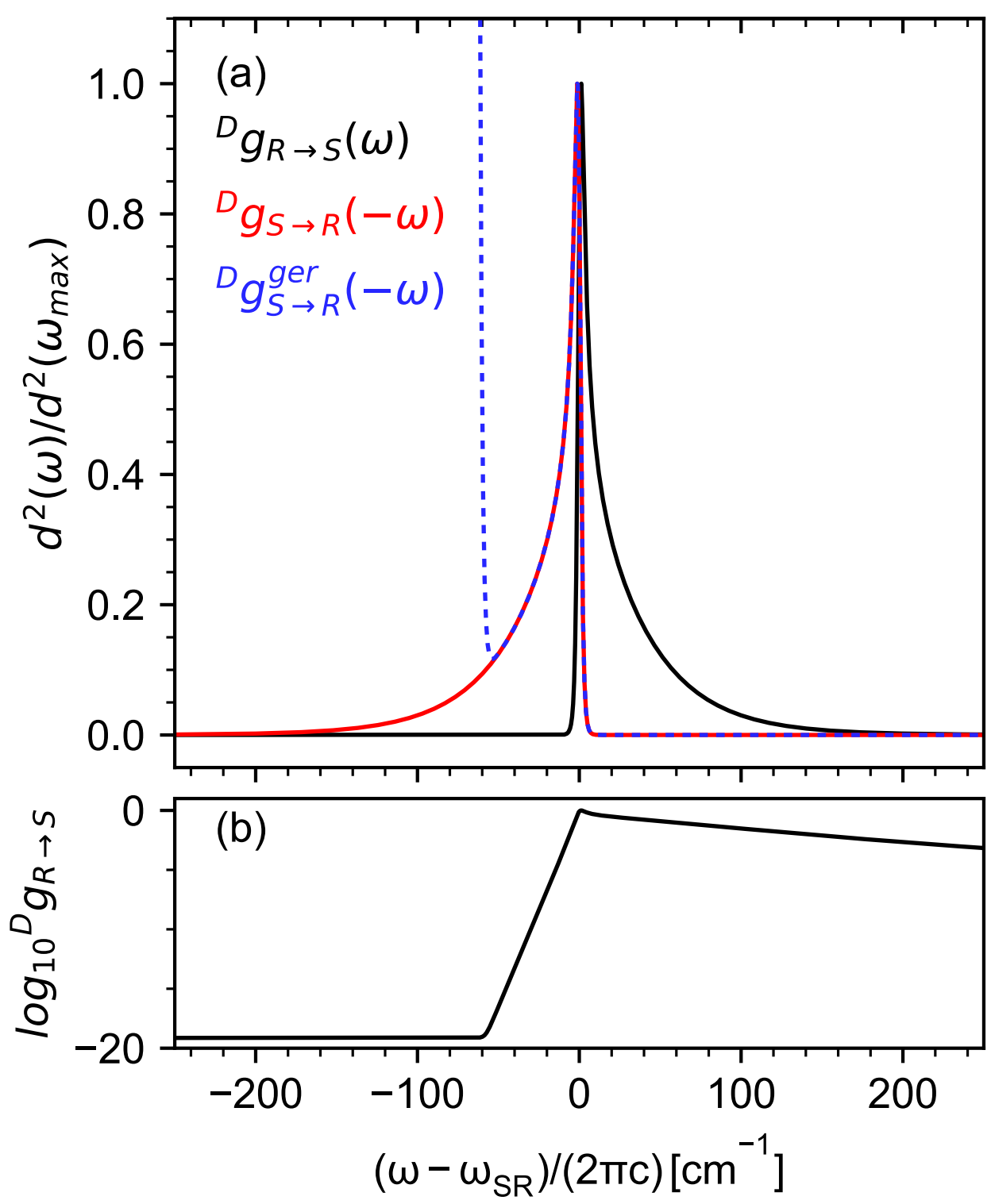

FIG. 6. Test of the quantum Brownian oscillator model against the generalized Einstein relation for a critically damped quantum oscillator with parameters from the FMO dimer model used in ref. 61, vibrational frequency $\omega_0/(2\pi c) = 70\ \mathrm{cm}^{-1}$ and reorganization energy $\lambda/(2\pi c) = 30\ \mathrm{cm}^{-1}$, at a temperature of 2K. A grid of 1048576-time steps with 0.1 fs spacing yielded a global precision of $10^{-17}$. (a) shows lineshapes for absorption (black) and stimulated emission (red) along with the GER-predicted stimulated emission lineshape (dashed blue). (b) shows the absorption lineshape on a logarithmic scale.

Figure 7 shows lineshapes for an under-damped quantum Brownian oscillator with $\omega_0/(2\pi c) = 1000\ \mathrm{cm}^{-1}$, $\gamma/(2\pi c) = 40\ \mathrm{cm}^{-1}$, and $\lambda/(2\pi c) = 1000\ \mathrm{cm}^{-1}$ at a temperature of 300K. These oscillator parameters correspond to those used in Fig. 2(b) of ref. 16. For this set of parameters, the reorganization energy is equal to a quantum of vibrational energy. These oscillator parameters approach the low-temperature limit since $k_{\mathrm{B}}T/(\hbar\omega_0) \approx 0.21 < 1$. The under-damped oscillator at a low temperature gives a series of resolved vibronic transitions in absorption and stimulated emission with maxima separated by the oscillator frequency, $\omega_0$. The Franck-Condon overlap factors[62] dictate the areas under these vibronic progressions. For example, the Franck-Condon overlap factors for these oscillator parameters predict that the 0-0 and 1-0 transitions have equal strengths, which agrees with the areas under those transitions. Figure 7(a) shows that the GER-predicted stimulated emission lineshape reproduces the low amplitude peaks in the stimulated emission lineshape at low frequencies. Even though the overlap between absorption and stimulated emission lineshapes is small at low frequencies, the generalized Einstein relation is obeyed. Again, the relative error in the GER-predicted stimulated emission [calculated by dividing the difference in Fig. 7(b) by the stimulated emission shown in Fig. 7(c)] is approximately equal to the relative error in the absorption lineshape [calculated by dividing the precision of absorption by the absorption shown in Fig. 7(c)]. The v=0 to v=1 transitions at $[(\omega - \omega_{SR}) = \pm\omega_0]/(2\pi c) \approx \pm 1000$ cm$^{-1}$ in Fig. 7 illustrate the relations between weak "hot-band" absorption from v" = 1 on *R* to v' = 0 on *S* and strong emission from v' = 0 on *S* to v" = 1 on *R* (both at –1000 cm$^{-1}$) and between weak "hot-band" emission from v' = 1 on *S* to v" = 0 on *R* and strong absorption from v" = 0 on *R* to v' = 1 on *S* (both at +1000 cm$^{-1}$). In principle, every transition in the low-frequency wing of the emission spectrum has a corresponding "hot band" in the absorption spectrum that is related to it by the exponential frequency and temperature dependence in the generalized Einstein relation of Eq. (7).

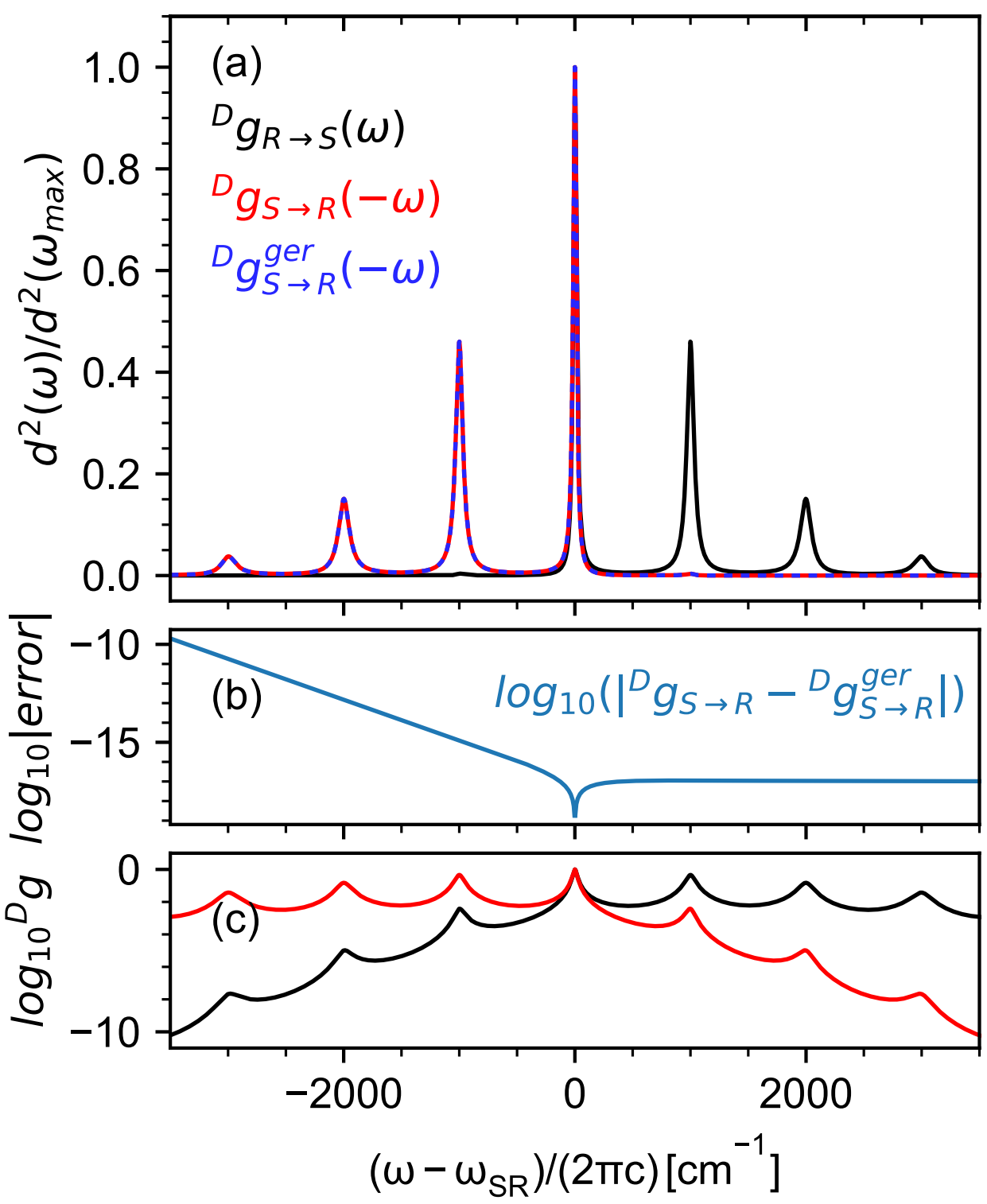

FIG. 7. Test of the quantum Brownian oscillator model against the generalized Einstein relation for an under-damped quantum oscillator with vibrational frequency $\omega_0/(2\pi c) = 1000\ \mathrm{cm}^{-1}$, damping constant $\gamma/(2\pi c) = 40\ \mathrm{cm}^{-1}$ and reorganization energy $\lambda/(2\pi c) = 1000\ \mathrm{cm}^{-1}$ at a temperature of 300K. A grid of 262144-time steps with 0.04 fs spacing yielded a global precision of $10^{-16}$. (a) shows lineshapes for absorption (black) and stimulated emission (red) along with the GER-predicted stimulated emission lineshape (dashed blue). (b) shows the absolute value of the difference between the stimulated emission lineshape and the GER-predicted stimulated emission lineshape on a logarithmic scale. (c) shows absorption and stimulated emission lineshapes on a logarithmic scale.

Mukamel's review discusses the Lorentzian limit of the semiclassical Brownian oscillator.[17] Figure 8 shows lineshapes for an over-damped quantum Brownian oscillator with $\omega_0/(2\pi c) = 1200\ \mathrm{cm}^{-1}$, $\gamma/(2\pi c) = 4800\ \mathrm{cm}^{-1}$, and $\lambda/(2\pi c) = 30.35\ \mathrm{cm}^{-1}$ at a temperature of 300K. The lineshapes are qualitatively near-Lorentzian, but with slightly asymmetric mirror-image tails. The Stokes' shift between absorption and stimulated emission is $\sim 5\ \mathrm{cm}^{-1}$. The GER-predicted stimulated emission lineshape visually agrees with the model's stimulated emission lineshape. The relative error in the GER-predicted stimulated emission [calculated by dividing the difference in Fig. 8(b) by the stimulated emission shown in Fig. 8(c)] is approximately equal to the relative error in the absorption lineshape [calculated by dividing the precision of absorption, $10^{-15}$, by the absorption shown in Fig. 8(c)]. Figure 8 shows that the quantum Brownian oscillator model can generate *approximately* Lorentzian lineshapes with similar widths and *practically negligible* Stokes' shifts that satisfy the generalized Einstein relation through skewed wings with absorption-emission mirror-image symmetry.

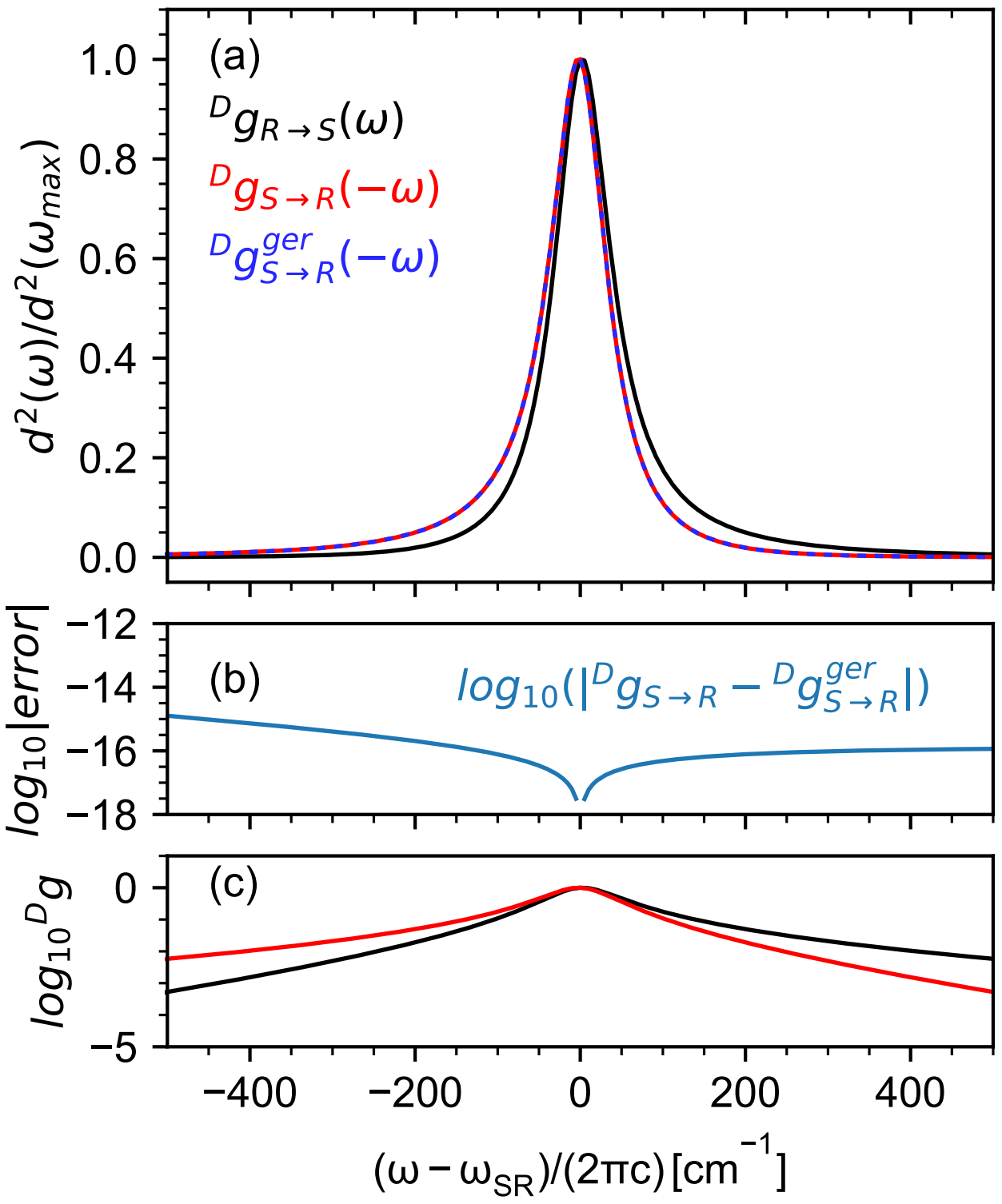

FIG. 8. Test of the quantum Brownian oscillator model against the generalized Einstein relation for an over-damped quantum Brownian oscillator with vibrational frequency $\omega_0/(2\pi c) = 1200\ \mathrm{cm}^{-1}$, damping constant $\gamma/(2\pi c) = 4800\ \mathrm{cm}^{-1}$ and reorganization energy $\lambda/(2\pi c) = 30.35\ \mathrm{cm}^{-1}$] at a temperature of 300K. A grid of 131072-time steps with 0.05 fs spacing yielded global precision of $10^{-15}$. (a) shows lineshapes for absorption (black) and stimulated emission (red) along with the GER-predicted stimulated emission lineshape (dashed blue). (b) shows the difference between the stimulated emission lineshape and the GER-predicted stimulated emission lineshape on a logarithmic scale. (c) shows absorption and stimulated emission lineshapes on a logarithmic scale.

Figure 9 shows the dipole-strength spectra and transition cross-sections for transitions between overlapping bands $R$ and $S$. The large damping frequency crudely models ultrafast damping of the solvated electron's coherent oscillations,[63] while the low frequency $\omega_{SR}$ is chosen to show the effect on transitions between overlapping bands that are roughly similar to the pre-solvated electron's infrared spectra.[64] Visually, the dipole-strength spectrum for the $R$ to $S$ transition is confined to positive frequencies, implying dominant absorption transitions. In contrast, the mirror-image dipole-strength spectrum for the $S$ to $R$ transition crosses zero frequency, showing that transitions from $S$ to $R$ contain some absorption in addition to their prevailing stimulated emission. Transition cross-sections are calculated from the dipole-strength spectra using Eq. (15). Each transition cross-section is an even function of frequency that is proportional to taking the difference between its dipole-strength spectrum and the mirror image of its dipole-strength spectrum and then multiplying by the frequency. This absolute frequency dependence of transition cross sections arises from radiation dynamics; unlike dipole-strength spectra, transition cross sections cannot be shifted by subtracting $\omega_{SR}$. The $R$ to $S$ transition cross-section in Fig. 9(b) is positive because this transition direction is dominantly absorptive. The $S$ to $R$ transition is mainly emissive with a negative transition cross-section in Fig. 9(b), but its mirror-image dipole-strength spectrum has some low-amplitude absorptive character below zero frequency in Fig. 9(a); this exactly cancels its dipole strength spectrum in the stimulated emission cross-section at zero frequency and partly cancels it nearby in Fig. 9(b). This absorption-emission cancellation near zero, combined with the $\omega$ frequency multiplier, makes each transition cross-section proportional to $\omega^2$ or a higher even power of the frequency near zero-frequency. Figure 9(a) shows that absorption-emission cancellation in the stimulated emission cross section is visually negligible near the emission maximum. The $\omega$ frequency multiplier in Eq. (15) is the dominant factor increasing the magnitude of the absorption cross-section relative to the stimulated emission cross-section by over a factor of 2.

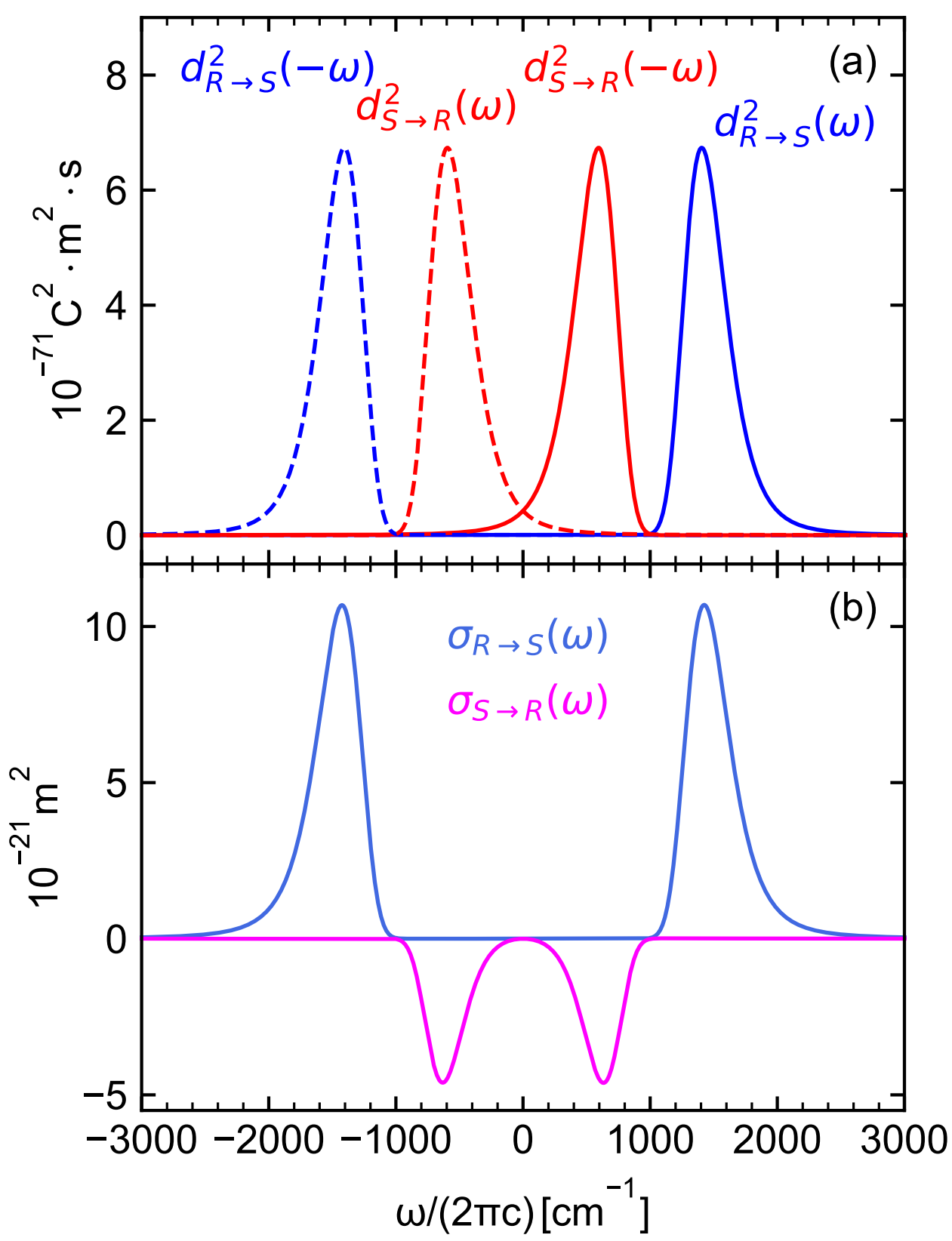

FIG. 9. Comparison of dipole-strength spectra and transition cross-sections between bands with 0-0 transition frequency $\omega_{SR}/(2\pi c) = 1000\ \mathrm{cm}^{-1}$ for an over-damped quantum oscillator with vibrational frequency $\omega_0/(2\pi c) = 500\ \mathrm{cm}^{-1}$, damping constant $\gamma/(2\pi c) = 6000\ \mathrm{cm}^{-1}$ and reorganization energy $\lambda/(2\pi c) = 500\ \mathrm{cm}^{-1}$ at a temperature of 20K. A grid of 262144-time steps with 0.1 fs spacing is used. The total dipole strength is $D^2_{R\to S} \approx 9.0 \times 10^{-58}\ \mathrm{C}^2 \cdot \mathrm{m}^2$, the refractive index is 1, and the local field factor is 1. (a) shows the dipole-strength spectrum for the $R$ to $S$ transition (solid blue curve), the mirror-image of the dipole-strength spectrum for the $S$ to $R$ transition (solid red curve), the mirror-image of the dipole-strength spectrum for the $R$ to $S$ transition (dashed blue curve), and the dipole-strength spectrum for the $S$ to $R$ transition (dashed red curve). (b) shows $R$ to $S$ (medium blue) and $S$ to $R$ (magenta) transition cross-sections.

## V. DISCUSSION

The generalized Einstein relations between absorption and emission rate processes were derived in refs. 24 and 27 by assuming detailed balance at equilibrium between single-photon transitions with Planck blackbody radiation and are valid when thermal quasi-equilibrium exists within each initial band.[24,59] Herein, the relationship between absorption and stimulated emission dipole-strength spectra was used to examine the internal consistency of Brownian oscillator lineshape models with detailed balance between single-photon absorption and stimulated emission transitions. In this section, we also discuss various other lineshape models and lineshape formulas in light of the generalized Einstein relations.

In the homogeneous Lorentzian lifetime broadening model,[1,30,31] the only source of line broadening is the population decay to the ground state, which occurs faster than the timescale to reach thermal quasi-equilibrium in the excited state. Since the lineshape model does not allow the system to achieve thermal quasi-equilibrium in the excited state, the generalized Einstein relations do not apply. The formula predicts the same Lorentzian lineshape for both absorption and emission spectra. Kubo's stochastic model[10,11] interpolates between the homogeneous Lorentzian and the inhomogeneous Gaussian limits, but implicitly assumes a high-temperature limit in which frequency fluctuations do not feel a thermal difference between up vs. down. This generates the same lineshape for absorption and stimulated emission and therefore fails to satisfy the generalized Einstein relations. Overdamped quantum Brownian oscillator models can generate lineshapes that are similar to those of the symmetric Lorentzian and stochastic model lineshapes[17] but obey the generalized Einstein relation through asymmetric wings. Absorption and emission lineshapes with Stokes' shift of 1 nm or less have been reported for 2D nanoplatelets of cadmium selenide (CdSe)[65–67] and perovskites,[68,69] and described as Lorentzian[67] or fit to Lorentzians.[66] Some of these narrow emission spectra (see Fig. 2 in ref. 65, Figs. 5 and 6 in ref. 68, or Fig. 1c in ref. 69) show clear wings with the skew found in Fig. 8. It would be interesting to examine whether these absorption and emission spectra obey the generalized Einstein relations, particularly for CdSe nanoplatelets with nanosecond lifetimes[65] that might be long enough for quasi-equilibrium emission to dominate the emission spectra.

If Doppler broadening is considered as inhomogeneous broadening, the generalized Einstein relations do not apply. The usual **k·v** treatment of Doppler broadening[70] leads to a Gaussian lineshape with photon-energy variance $\Delta^2 = (\hbar\omega_{sr})^2 k_{\mathrm{B}}T/M_0c^2$ where $\omega_{sr}$ is the transition frequency and $M_0$ is the rest mass of the molecule. However, if the excited state is prepared at thermal equilibrium or lives long enough for the molecule to attain a thermal equilibrium velocity distribution in the excited state, the lineshape becomes homogeneous and the generalized Einstein relations apply. For a homogeneous Gaussian absorption lineshape, the generalized Einstein relation in Eq. (7) predicts that the stimulated emission lineshape is also a Gaussian with the same variance, and that the variance is related to the photon-energy Stokes' shift ($2\lambda$) by $\Delta^2 = (2\lambda)k_{\mathrm{B}}T$.[59] The supplementary material shows that this Stokes' shift matches the photon recoil shift energy $|\hbar\mathbf{k}_{sr}|^2/M_0$ calculated with non-relativistic mechanics and constant photon momentum $\hbar\mathbf{k}_{sr}$ over the linewidth,[71] a point made by McCumber.[72] However, such a rough treatment of Doppler broadening does not consistently incorporate collisional relaxation, and Dicke has shown that collisions could lead to narrowing of Doppler broadened spectra.[73] Ryu *et al.* showed that under thermal equilibrium with Planck blackbody radiation, purely radiative lifetime broadening is compatible with the generalized Einstein relations.[24] The essential aspect is that blackbody radiation preferentially excites the low energy side of the lineshape.

The semi-classical Brownian oscillator model couples the quantum molecular vibration to a classical bath. It is known that classical bath approaches can lead to violations of detailed balance in the quantum sub-system even when the bath is adequately described by classical mechanics.[74] Since the model does not obey the generalized Einstein relation, Eq. (7), it does not satisfy detailed balance between absorption and stimulated emission. The model's departures from the generalized Einstein relation, such as the 3% departure near the emission maximum in Fig. 3(a), could be experimentally detectable, so the semi-classical model is not a useful guide for designing experimental tests of whether the relations hold for a given sample. More strangely, absorption lineshapes calculated using the model also show smaller but numerically significant negative features on their red-edges. The quantum Golden Rule gives non-negative dipole-strength spectra,[27] so these negative regions are problematic.[56,75,76] It is not clear whether such unphysical negative regions are specific to the model (classical Langevin force or lack of bath memory)[15] or may be more generally problematic for treatments of a single lineshapes using a quantum system in a classical bath.

The quantum Brownian oscillator model developed by Caldeira and Leggett provides an exact treatment of harmonic vibrational motion that is bi-linearly coupled to a thermal bath of quantum harmonic oscillators.[35] Tanimura and Mukamel[15] and Gu, Widom and Champion,[16] extended this model to treat transitions between two displaced, but otherwise identical harmonic potentials with the same Brownian oscillator on each state and the same coupling to the same thermal bath of quantum harmonic oscillators for both states. Every converged lineshape we have calculated using this two-state quantum Brownian oscillator model provides absorption and stimulated emission spectra that obey the generalized Einstein relation to within numerical precision. This suggests that the equilibrium quantum Brownian oscillator lineshapes obey the generalized Einstein relation. Among all the lineshape models discussed in this work, only the quantum Brownian oscillator model satisfies detailed balance between absorption and stimulated emission. However, the model does not include relaxation between bands and does not include spontaneous emission transitions (see discussions below Eq. (4.65) and below Eq. (9.13) in ref. 18). Spontaneous emission could be phenomenologically added to the model by invoking the generalized Einstein relationship between stimulated emission and spontaneous emission lineshapes in Eq. (5b). However, such addition of a new decay pathway leads to additional lineshape broadening that is not accounted for by the model's harmonic oscillator bath. To account for this additional broadening, one would have to either effectively reinterpret the quantum harmonic oscillator bath or treat additional lineshape broadening through interactions with the quantum radiation field. With more significant spontaneous emission rate contributions to the linewidth, the physical interpretation of the effective quantum Brownian oscillator model parameters that fit the lineshape becomes less accurate.

Quantum Brownian oscillator model lineshapes always have a Stokes' shift and mirror-image symmetry between absorption and stimulated emission. This symmetry arises from the harmonic model and is not required by the generalized Einstein relations. To generate phenomenological asymmetric lineshapes that obey the generalized Einstein relations, quantum Brownian oscillator lineshapes can be attached to quantum levels as outlined for any pair of lineshapes obeying the generalized Einstein relation in Appendix B of ref. 27. Such calculations could be used to model the absorption and emission spectra of fairly rigid conjugated molecules in solution that reach quasi-equilibrium in picoseconds[77] and have near-radiative nanosecond lifetimes,[78] so that quasi-equilibrium emission dominates. Such model spectra could predict the experimental accuracy needed for tests of the generalized Einstein relations.

The relationship between dipole strength spectra and transition cross-sections developed in ref. 27 and applied to lineshapes from time-correlation functions here gives *individual* transition cross-sections from the quantum Brownian oscillator model the correct dipolar low-frequency behavior. Such quadratic low-frequency behavior is an experimentally verified aspect of the equilibrium Debye,[79] Van Vleck-Weisskopf,[4] and Ben-Reuven[6] lineshapes. At equilibrium, the relationship developed here groups absorption and stimulated emission terms into signed cross-sections differently from previous correlation function treatments with positive cross-sections.[27,75] The grouping in prior correlation-function treatments gives equilibrium net absorption (absorption minus stimulated emission) in the rotating-wave approximation [see the absorption cross-section in Eq. (II.1) of ref. 75 or the absorption coefficient obtained by substituting Eq. (6.9b) into Eq. (4.64) in ref. 18]. With the re-grouping, the rotating wave approximation is repaired by combining the two terms in Eq. (15), net absorption is calculated from forward and reverse transition cross-sections with independent initial band populations in Eq. (14), and the required quadratic (or higher even power) low-frequency behavior occurs for every transition cross-section at quasi-equilibrium (as illustrated in Fig. 9). In consequence, the treatment here correctly predicts the required overall quadratic (or higher even power) low-frequency stimulated transition behavior even for transitions with spectra that cross zero-frequency and systems with non-equilibrium band populations.

## VI. CONCLUSIONS

We have investigated Brownian oscillator lineshape models for their compatibility with the generalized Einstein relation between absorption and stimulated emission dipole-strength spectra, which is derived assuming detailed balance between single-photon transitions and Planck blackbody radiation. The Bloch model, the stochastic model, and the semi-classical Brownian oscillator model do not obey the generalized Einstein relation and therefore fail to satisfy detailed balance. The semi-classical Brownian oscillator model has negative regions in the optical absorption lineshape. Spectra calculated using the two-state quantum Brownian oscillator model obey the generalized Einstein relation within the numerical precision of the calculation, suggesting that the lineshape model satisfies detailed balance between absorption and stimulated emission. The relationship between lineshapes obtained via Fourier transformation of time-correlation functions and stimulated transition cross-sections was developed in a form suitable for transitions so broad that they are not uniquely classifiable as absorption or emission and situations in which bands are each internally equilibrated, but band populations are not equilibrated with each other.

## SUPPLEMENTARY MATERIALS

See supplementary materials for the demonstration of FFT aliasing wrap-around error, effect of the Matsubara sum on the absorption lineshape, accuracy of the FFT calculations using homogeneous Gaussian lineshapes, semi-classical Brownian oscillator lineshapes for thermal energy greater than vibrational energy, generalized Einstein relation tests on quantum Brownian oscillator lineshapes at different temperatures, and a demonstration that equilibrium Doppler broadening satisfies the generalized Einstein relation. Data and Fortran 95/2003 functions for the Matsubara sum are available as separate files.

## ACKNOWLEDGEMENTS

This material is based upon work supported by the National Science Foundation under award number CHE-2155010.

Conflict of Interest Statement – The authors have no conflicts to disclose.

Author Contributions – D.M.J. designed research, J.R. rewrote functions for the Matsubara sums and did initial lineshape calculations, A.K.A. calculated lineshapes and transition cross-sections, A.K.A. wrote the initial draft with rewriting mainly by A.K.A. and D.M.J. and revisions by all authors.

Data Availability – The data that support the findings of this study are available within the article and its supplementary material.

# Supplementary Materials

for

# Optical Lineshape Models and the Generalized Einstein Relation between Absorption and Stimulated Emission

Aman K. Agrawal, Jisu Ryu, and David M. Jonas

Department of Chemistry, University of Colorado, Boulder, CO, 80309, USA.

**The file includes:**

**Figures S1 to S7**

**Photon Recoil, Doppler Broadening and the Generalized Einstein Relations**

**References**

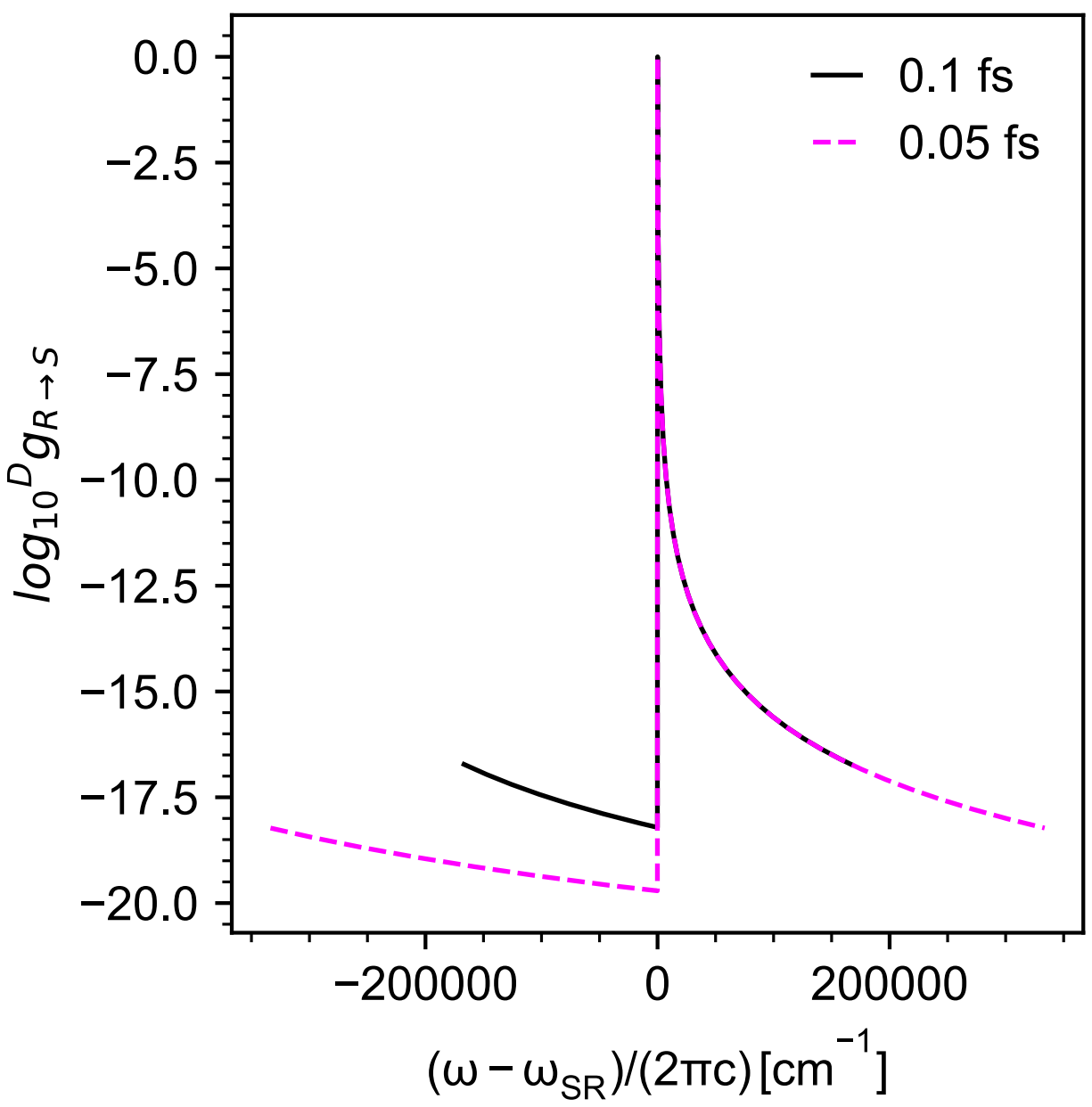

FIG. S1. Demonstration of FFT aliasing wrap-around error on the absorption lineshape for the over-damped quantum Brownian oscillator of Fig. 2 [vibrational frequency $\omega_0/(2\pi c) = 50\ \mathrm{cm}^{-1}$, damping constant $\gamma/(2\pi c) = 600\ \mathrm{cm}^{-1}$ and reorganization energy $\lambda/(2\pi c) = 50\ \mathrm{cm}^{-1}$] at the same temperature of 4K. Black solid (magenta dashed) curve is the height-normalized absorption lineshape calculated on a grid of 131072- (262144-) time steps with 0.1 (0.05) fs spacing.

**Alt Text:**

**Short:** Line graph depicting dipole-strength spectra for absorption against frequency calculated for two different time step sizes.

**Long:** A graph depicting dipole strength spectra with the horizontal axis labelled omega minus omega subscript SR divided by two pi c and the vertical axis labelled log base 10 of superscript D g subscript R rightarrow S. The graph features two curves: a solid black curve labelled 0 point 1 fs, and a dashed magenta curve labelled 0 point 05 fs. The black curve is in the frequency range from about negative 160000 to positive 160000 wavenumbers, while the magenta curve is in the frequency range from about negative 320000 to positive 320000 wavenumbers. The black curve initially decreases from negative 16.5 to negative 18 at around zero frequency, increases steeply to 0 around zero frequency and then slowly decays to negative 16.5. The magenta curve initially decreases from negative 18 to negative 20 at around zero frequency, increases steeply to 0 around zero frequency and then slowly decays to negative 18. The two curves overlap for positive frequencies in their common range.

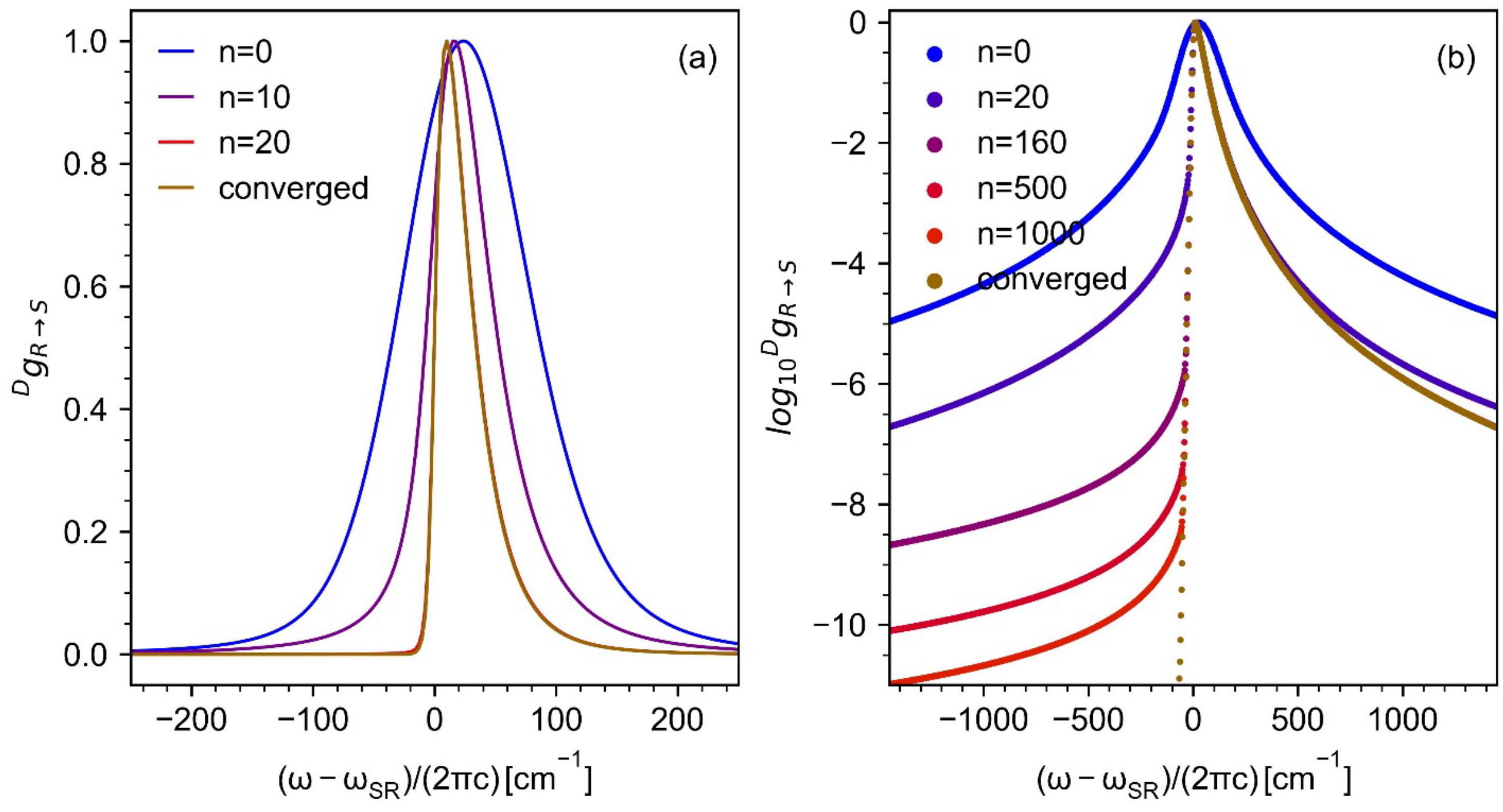

FIG. S2. Effect of the number of terms in the Matsubara sum on the absorption lineshape for an over-damped quantum Brownian oscillator with vibrational frequency $\omega_0/(2\pi c) = 70\ \mathrm{cm}^{-1}$, damping constant $\gamma/(2\pi c) = 280\ \mathrm{cm}^{-1}$ and reorganization energy $\lambda/(2\pi c) = 30\ \mathrm{cm}^{-1}$ at a temperature of 4K. The converged calculation included 19169900 terms. A grid of 131072-time steps with 0.1 fs spacing yielded lineshape global precision of $10^{-11}$ for the converged absorption lineshape. (a) shows absorption lineshapes and (b) shows absorption lineshapes on a logarithmic scale.

**Alt Text:**

**Short:** 2 panels with line graphs depicting absorption dipole-strength spectra against frequency for different numbers of terms in the Matsubara sum.

**Long:** The figure consists of two panels, labeled a and b. Both panels have graphs with the x-axis labelled omega minus omega subscript SR divided by two pi c. The graph in panel a has y-axis labelled left superscript D g subscript R rightarrow S, while the graph in panel b has y-axis labelled log base 10 of superscript D g subscript R rightarrow S. In panel a, four colored curves representing absorption lineshapes are labeled with the number of terms in the Matsubara sum: 0 in blue, 10 in purple, 20 in red, and the term "converged" in brown. All curves approach zero on either side of the peak, which is to the right of zero frequency. The curve with zero terms is symmetric around the peak, while the curves get more asymmetric with a maximum to the left of the peak and a tail to the right of the peak as the number of terms increases. The red and brown curves overlap except near the baseline below zero frequency, where the brown curve is lower..

Panel b has an x-axis ranging from negative 1450 to positive 1450 wavenumbers and a log base 10 y-axis ranging from minus 11 to 0. It shows six colored scatter plots representing absorption lineshapes for different numbers of terms in the Matsubara sum: 0 in blue, 20 in deep blue, 160 in purple, 500 in red, 1000 in orange and the term "converged" in brown. Except for 0 and 20, which lie above the others, the curves match for positive frequencies. For negative frequencies, the curves with more terms in the Matsubara sum show more negative logarithms. Starting from the lowest frequency, the logarithms all monotonically increase towards the lineshape maximum. The converged curve drops below the lower y-axis limit of negative 11 around negative 80 wavenumbers.

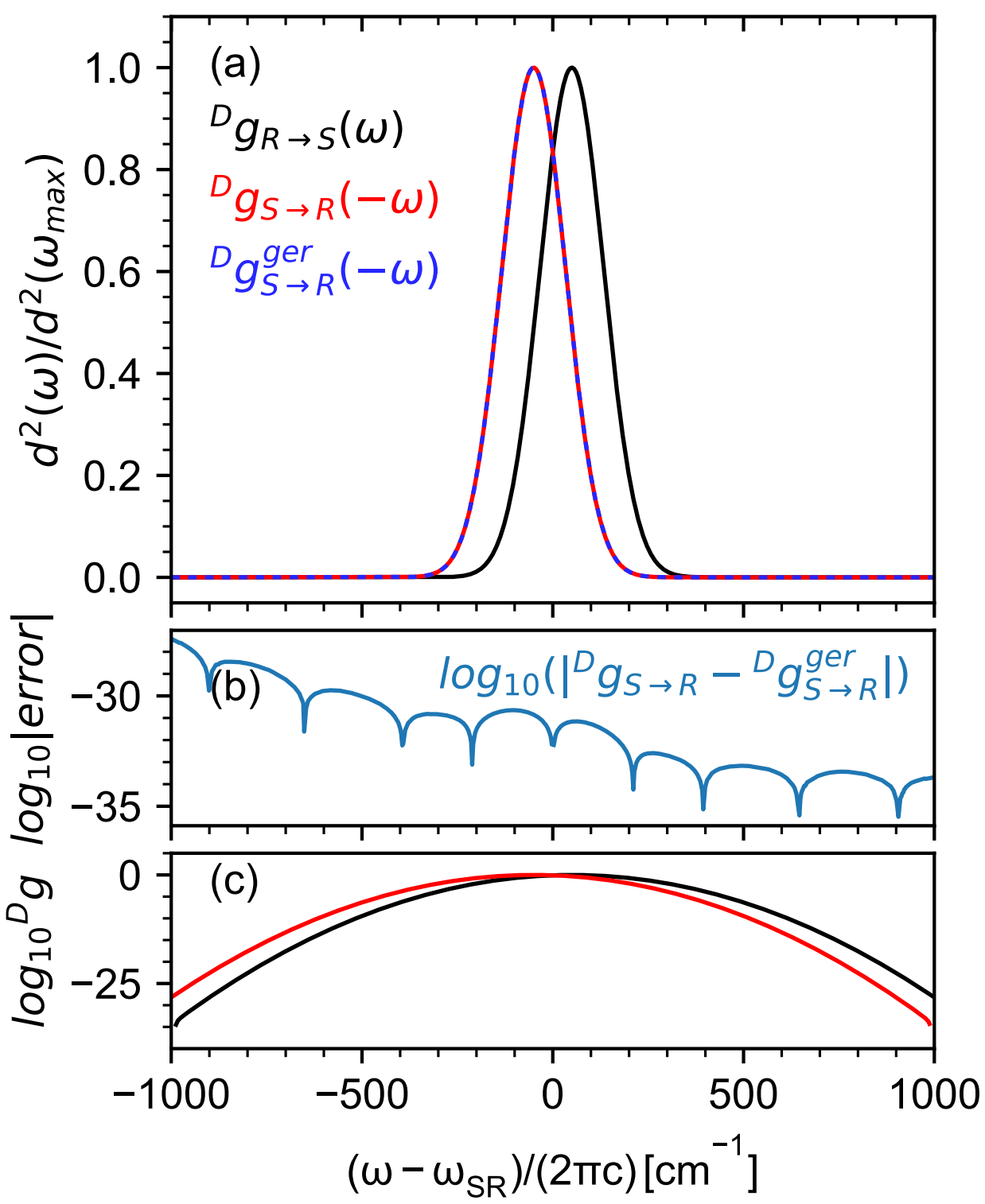

FIG. S3. A check on the accuracy of the FFT calculations using the homogeneous Gaussian lineshape that analytically follows the generalized Einstein relation. (a) shows the dipole-strength lineshape for absorption (black) and stimulated emission (red) calculated from the line-broadening formula with reorganization energy $\lambda/(2\pi c) = 50\ \mathrm{cm}^{-1}$ at a temperature of 100K. The stimulated emission lineshape (dashed blue) is calculated from the absorption lineshape using the generalized Einstein relation. A grid of 131072-time steps with 0.1 fs spacing was used for the calculation. (b) shows the difference between the stimulated emission lineshape obtained using the model and the GER-predicted stimulated emission lineshape on a logarithmic scale. (c) shows FFT-calculated lineshapes on a logarithmic scale.

**Alt Text:**

**Short:** 3 panels with line graphs depicting dipole-strength spectra against frequency and the generalized Einstein relation prediction of emission from absorption for Gaussian lineshapes.

**Long:** The figure consists of 3 panels, labeled (a), (b), and (c). All panels have the same x-axis labelled omega minus omega subscript SR divided by two pi c and spanning negative 1000 to positive 1000 wavenumbers. In panel (a), the vertical axis is labelled d superscript 2 as function of omega divided by the maximum of d superscript 2. Panel (a) features three curves: a solid black curve labelled superscript D g subscript R rightarrow S as a function of omega; a solid red curve labelled superscript D g subscript S rightarrow R as a function of negative omega; and a dashed blue curve labelled superscript D g superscript ger, subscript S rightarrow R as a function of negative omega. The black

Gaussian curve peaks to the right of zero at positive 50 wavenumbers, the red Gaussian curve peaks to the left of zero at negative 50 wavenumbers and is the mirror image of the black curve across zero, and the blue dashed curve overlaps the red curve.

Panel (b) has vertical axis labelled log base 10 of absolute error and shows a light blue solid curve labelled log base 10 of absolute value of difference between superscript D g subscript S rightarrow R and superscript D g superscript ger, subscript S rightarrow R. The curve has 9 regularly spaced cusps extending below a general decrease from negative 27 to negative 35 as the x-axis goes from negative 1000 to positive 1000.

Panel (c) has vertical axis labelled log base 10 of superscript D g and shows a black solid curve and a red solid curve. The black curve is a downward parabola that peaks at 0 to the right of zero frequency at positive 50 wavenumbers, the red curve is a downward parabola that peaks at 0 to the left of zero at negative 50 wavenumbers. The black curve increases from negative 35 to 0 and then decreases to about negative 28, while the red curve increases from negative 28 to 0 and then decreases to about negative 35. The two curves cross at zero frequency.

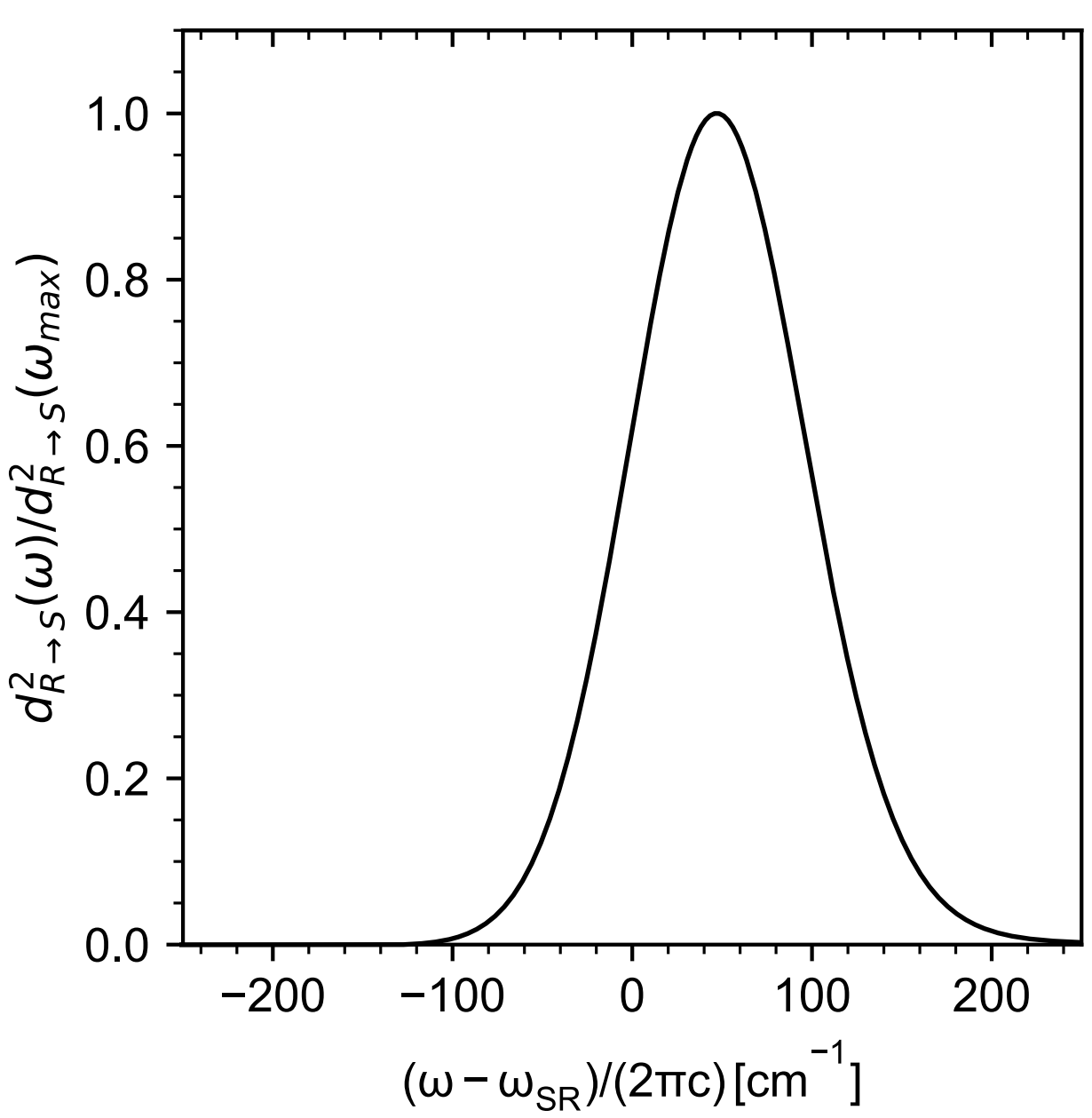

FIG. S4. Absorption lineshape calculated using the semi-classical Brownian oscillator model of ref. 1 for an over-damped oscillator with same parameters as used in Fig. 3 here and in Fig. 1 of ref. 2 [vibrational frequency $\omega_0/(2\pi c) = 50\ \mathrm{cm}^{-1}$, damping constant $\gamma/(2\pi c) = 600\ \mathrm{cm}^{-1}$ and reorganization energy $\lambda/(2\pi c) = 50\ \mathrm{cm}^{-1}$], at the same temperature of 2K as in Fig. 1a of ref. 2. A grid of 131072-time steps with 0.1 fs spacing yielded global precision of $10^{-15}$.

**Alt Text:**

**Short:** Line graph depicting dipole-strength spectra for absorption against frequency.

**Long:** A graph depicting dipole strength spectra for absorption with the x-axis labelled omega minus omega subscript SR divided by two pi c and ranging from negative 250 to positive 250 wavenumbers. The vertical axis is labelled d superscript 2, subscript R rightarrow S as function of omega divided by the maximum of d superscript 2, subscript R rightarrow S and ranges from 0.0 to 1.1. The graph features a solid black curve with an approximately Gaussian shape that peaks with a height of 1 between 40 and 60 wavenumbers.

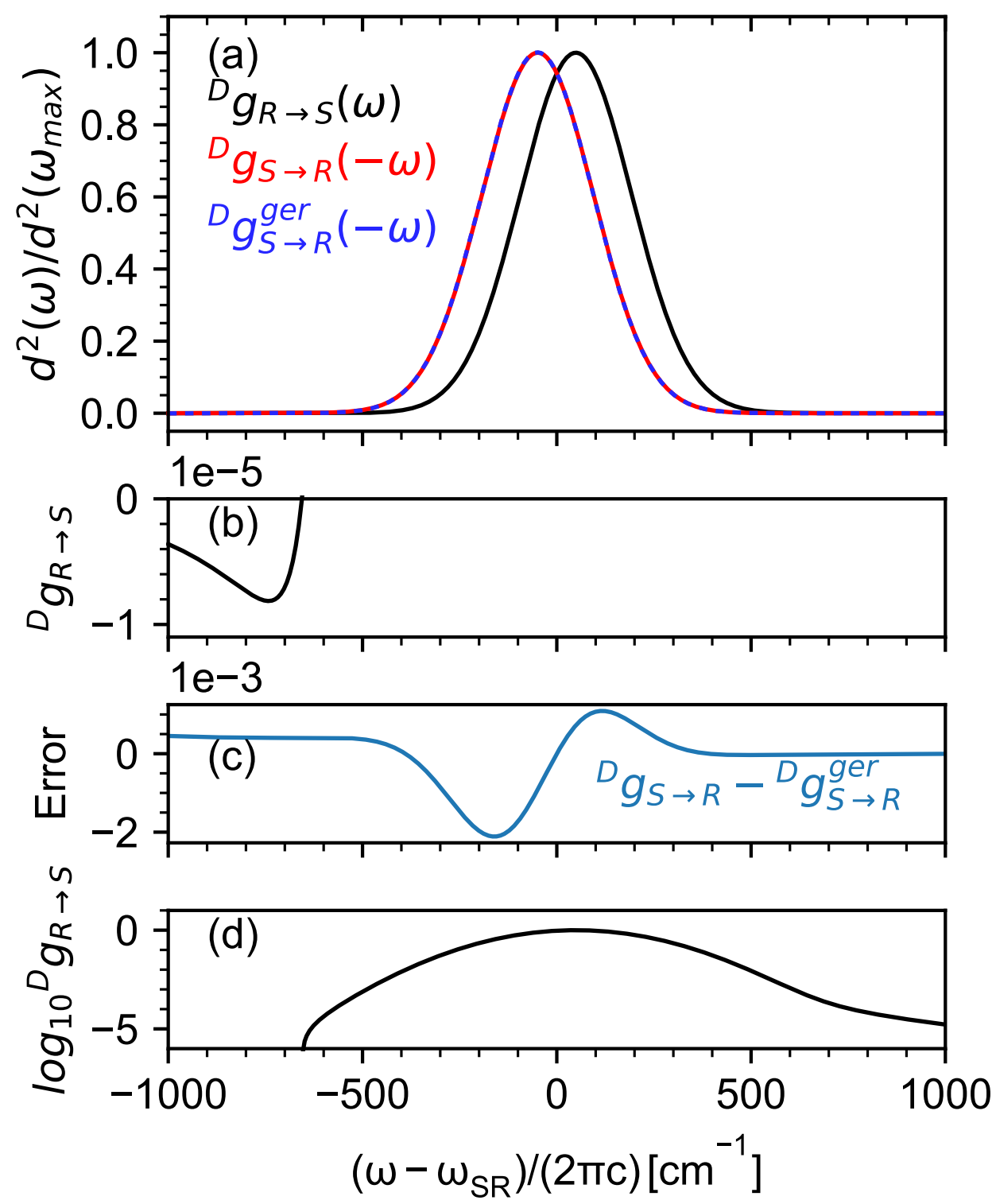

FIG. S5. Test of the semi-classical Brownian oscillator model of ref. 1 against the generalized Einstein relation for an over-damped oscillator with the same parameters as used in Fig. 3 here and in Fig. 1 of ref. 2 [vibrational frequency $\omega_0/(2\pi c) = 50\ \mathrm{cm}^{-1}$, damping constant $\gamma/(2\pi c) = 600\ \mathrm{cm}^{-1}$ and reorganization energy $\lambda/(2\pi c) = 50\ \mathrm{cm}^{-1}$], at the same temperature of 300K as in Fig. 1b of ref. 2. A grid of 131072-time steps with 0.1 fs spacing yielded global precision of $10^{-15}$. (a) shows lineshapes for absorption (black) and stimulated emission (red) along with the GER-predicted stimulated emission lineshape (dashed blue). (b) shows the negative part of the absorption lineshape. (c) shows the difference between the stimulated emission lineshape obtained using the model and the GER-predicted stimulated emission lineshape. (d) shows the positive region of the absorption lineshape on a logarithmic scale.

**Alt Text:**

**Short:** 4 stacked panels with line graphs depicting dipole-strength spectra against frequency and the generalized Einstein relation prediction of emission from absorption.

**Long:** The figure consists of 4 stacked panels, labeled (a), (b), (c), and (d). All panels have the same x-axis labelled omega minus omega subscript SR divided by two pi c and ranging from negative 1000 to positive 1000 wavenumbers. In (a) the vertical axis is labelled d superscript 2 as function of omega divided by the maximum of d superscript 2 and ranges from 0 to 1. Panel (a) features three curves that all look roughly Gaussian in shape: a solid black curve labelled superscript D g subscript R rightarrow S as a function of omega; a solid red curve labelled superscript D g subscript S rightarrow R as a function of negative omega; and a dashed blue curve labelled superscript D g superscript ger, subscript S rightarrow R as a function of negative omega. The black curve peaks at 1 to the right of zero near positive 50 wavenumbers and the red curve peaks at 1 to the left of zero near negative 50 wavenumbers. The blue dashed curve matches the red curve.

Panel (b) has vertical axis labelled superscript D g subscript R rightarrow S and shows a black solid curve from panel a. The vertical axis is enlarged to go from negative 1 times $10^{-5}$ to zero. The curve is negative 4 times $10^{-6}$ at negative 1000 wavenumbers and dips to negative 8 times $10^{-6}$ at negative 750 wavenumbers, then quickly rises above zero at negative 650 wavenumbers.

Panel (c) has vertical axis labelled error and shows a light blue solid curve labelled superscript D g subscript S rightarrow R minus superscript D g superscript ger, subscript S rightarrow R. The curve is almost flat at 5 times $10^{-4}$ before decreasing to negative 2 times $10^{-3}$ at negative 150 wavenumbers. The curve rises through zero at zero frequency to 1 times $10^{-3}$ at 100 wavenumbers, then decreases to almost zero at 400 wavenumbers and remains flat. The shape of the curve resembles the negative of the derivative of a Gaussian, but the maximum on the right has smaller magnitude than the minimum on the left.

Panel (d) has vertical axis labelled log base 10 of superscript D g subscript R rightarrow S that ranged from negative 6 to 0. The black solid curve peaks to the right of zero frequency at 0 on the vertical axis and decreases to the right and left of the peak. The curve heads straight down and abruptly ends at around negative 650 wavenumbers on the left-hand side.

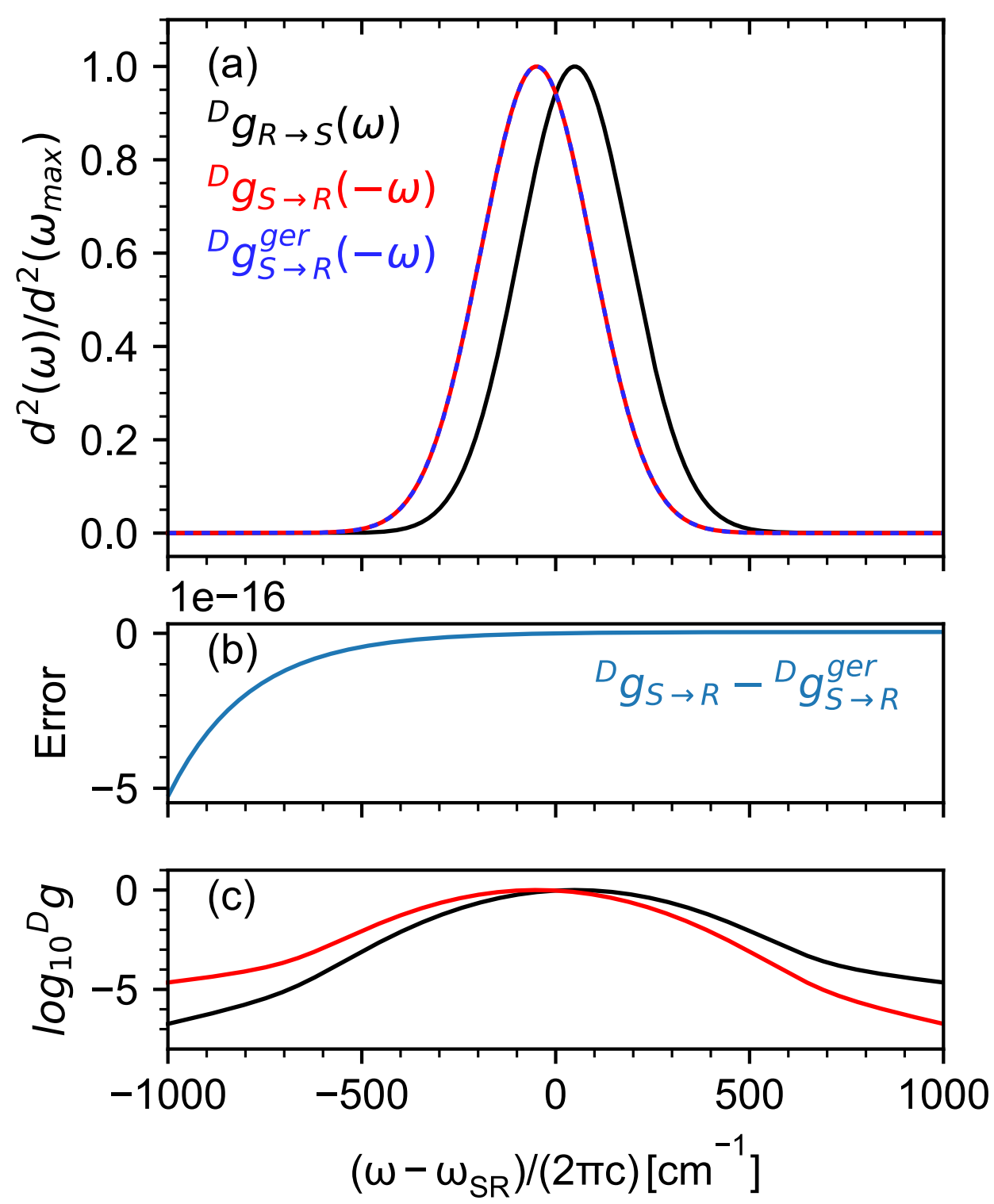

FIG. S6. Test of the quantum Brownian oscillator model of ref. 2 against the generalized Einstein relation for the over-damped oscillator of Fig. 4 here and Fig. 1 of ref. 2 [vibrational frequency $\omega_0/(2\pi c) = 50\ \mathrm{cm}^{-1}$, damping constant $\gamma/(2\pi c) = 600\ \mathrm{cm}^{-1}$ and reorganization energy $\lambda/(2\pi c) = 50\ \mathrm{cm}^{-1}$], at the same temperature of 300K as in Fig. 1b of ref. 2. A grid of 131072-time steps with 0.1 fs spacing yielded global precision of $10^{-17}$. (a) shows lineshapes for absorption (black) and stimulated emission (red) along with the GER-predicted stimulated emission lineshape (dashed blue). (b) shows the difference between the stimulated emission lineshape obtained using the model and the GER-predicted stimulated emission lineshape. (c) shows lineshapes on a logarithmic scale.

**Alt Text:**

**Short:** 3 stacked panels with line graphs depicting dipole-strength spectra against frequency and the generalized Einstein relation prediction of emission from absorption.

**Long:** The figure consists of 3 panels, labeled (a), (b), and (c). All panels have the same x-axis labelled omega minus omega subscript SR divided by two pi c and ranging from negative 1000 to positive 1000 wavenumbers. In (a) the vertical axis is labelled d superscript 2 as function of omega divided by the maximum of d superscript 2 and ranges from 0 to 1. Panel (a) features three roughly Gaussian curves: a solid black curve labelled superscript D g subscript R rightarrow S as a function of omega; a solid red curve labelled superscript D g subscript S rightarrow R as a function of negative omega; and a dashed blue curve labelled superscript D g superscript ger, subscript S rightarrow R as a function of negative omega. The black curve peaks at a height of 1 near positive 50 wavenumbers, the red curve peaks at a height of 1 near negative 50 wavenumbers, and the blue dashed curve overlaps the red curve.

Panel (b) has vertical axis labelled error that ranges from a minimum of negative 5 times ten to the minus 16 to a maximum of zero and shows a light blue solid curve labelled superscript D g subscript S rightarrow R minus superscript D g superscript ger, subscript S rightarrow R. The curve rises from negative 5 times $10^{-16}$ at negative 1000 wavenumbers to almost zero at negative 100 wavenumbers and then remains flat at zero.

Panel (c) has vertical axis labelled log base 10 of superscript D g that ranges from negative 8 to 0 and shows a black solid curve and a red solid curve. The black curve peaks to the right of zero, the red curve peaks to the left of zero and both curves decrease on either side of their peaks. The black curve increases from negative 7 to 0 and then decreases to about negative 5, while the red curve increases from negative 5 to 0 and then decreases to about negative 7. Both curves are approximately parabolic downward near their maximum but then transition to slower, roughly linear decays on both sides of their maximum at around plus minus 750 wavenumbers.

## Photon Recoil, Doppler Broadening and the Generalized Einstein Relations

McCumber noted that Doppler broadening satisfies the generalized Einstein relation between absorption and stimulated emission when the photon recoil energy is taken into account.[3] Here, Doppler broadening and photon recoil[4] are treated by considering conservation of momentum and energy during the radiative transition. Doppler broadening follows the generalized Einstein relation if the velocity distribution in the excited state starts at or reaches thermal quasi-equilibrium without the inhomogeneous Doppler lineshape changing through collisions. Consider the inhomogeneously Doppler broadened transition between two levels separated by energy $\hbar\omega_{sr}$ for a molecule with rest mass $M_0$ at temperature $T$. Assuming non-relativistic mechanics and approximating the photon momentum as constant over the linewidth ($\hbar k \approx \hbar k_{sr}$), Fig. S7 shows the parabolic dependence ($p^2/2M_0$) of the internal + 1D translational kinetic energy for each band as a function of the projection of the molecular momentum in the lower band *r* along the photon momentum. The upper band $s$ is shifted sideways from the lower band *r* by the photon momentum, $\hbar k_{sr}$, where $k_{sr} = \omega_{sr}/c$ (*c* is the speed of light in vacuum) is the wavevector for the photon with frequency $\omega_{sr}$.

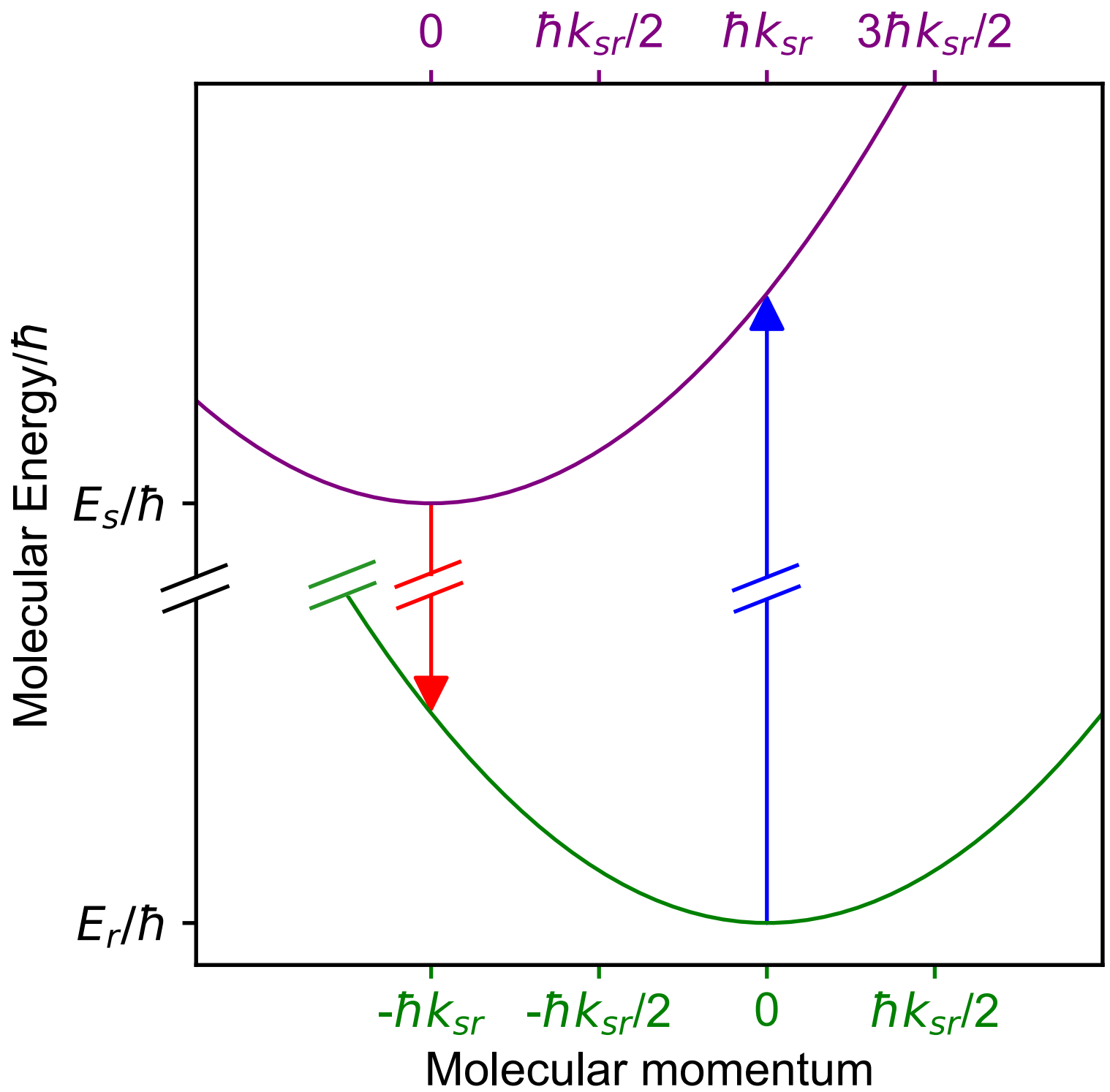

FIG. S7. Band diagram to illustrate Doppler broadening of the absorption and emission lineshapes. The green lower curve and the purple upper curve show the parabolic dependence of the total energy for lower band *r* and upper band *s*, respectively, as a function of the molecular translational momentum. The lower x-axis is the molecular momentum for the lower band, and the upper x-axis is the molecular momentum for the upper band. The blue (red) vertical line shows the absorption (emission) from the lower (upper) band to the upper (lower) band for zero molecular momentum in the initial band.

**Alt Text:**

**Short:** Two parabolic total energy curves as a function of molecular translational momentum with absorption and emission photon energies depicted as vertical arrows.

**Long:** A graph showing single-photon transitions with upper and lower horizontal axes labeled as molecular momentum and the left vertical axes labeled as molecular energy divided by hbar. A solid green parabolic curve upward represents the total energy of the lower band italic r and has its minimum at momentum 0 on the green lower horizontal axis, momentum hbar times k subscript sr on the purple upper horizontal axis, and energy E subscript r over hbar on the vertical axis A solid purple parabolic curve upward represents the total energy of the upper band italic s and has its minimum at momentum minus hbar times k subscript sr on the green lower horizontal axis, at momentum 0 on the purple upper horizontal axis, and energy E subscript s over hbar. The blue vertical arrow goes from the lower green curve at zero momentum on the green lower horizontal axis to the upper purple curve at momentum of hbar times k subscript sr on the purple upper horizontal axis, while the red vertical arrow goes from the purple upper curve at zero momentum on the purple upper horizontal axis to the green lower curve at a momentum of negative hbar times k subscript sr on the green lower horizontal axis.

For transitions between lower band *r* and upper band *s* for a molecule with momentum $p$ in the lower band, the absorption frequency is

$$\omega(p) = \omega_{sr} + \frac{(p + \hbar k_{sr})^2}{2M_0\hbar} - \frac{p^2}{2M_0\hbar} = \omega_{sr} + \frac{\hbar k_{sr}^2}{2M_0} + \frac{pk_{sr}}{M_0}, \qquad (S1a)$$

and the (negative) emission frequency for a molecule with momentum $p$ in the upper band is

$$-\omega(p) = \omega_{sr} + \frac{p^2}{2M_0\hbar} - \frac{(p - \hbar k_{sr})^2}{2M_0\hbar} = \omega_{sr} - \frac{\hbar k_{sr}^2}{2M_0} + \frac{pk_{sr}}{M_0}. \qquad (S1b)$$

Within each band, the equilibrium Maxwell-Boltzmann probability distribution for momentum is

$$P^p(p) = \frac{1}{\sqrt{2\pi M_0 k_\mathrm{B} T}} \exp\left(-\frac{p^2}{2M_0 k_\mathrm{B} T}\right). \qquad (S2)$$

The inhomogeneous equilibrium absorption and emission lineshapes can then be calculated by inverting Eqs. (S1a) and (S1b) to obtain $p(\omega)$ and substituting into Eq. (S2).

$$g_{\substack{abs \\ em}}(\pm\omega) = g_{max} \exp\left(-\frac{M_0\left(\omega - \left(\omega_{sr} \pm \frac{\hbar k_{sr}^2}{2M_0}\right)\right)^2}{2k_{sr}^2 k_B T}\right). \qquad (S3)$$

The change in standard chemical potential is $\Delta\mu^\circ_{r\to s} = \hbar\omega_{sr}$. Equation (S3) shows that the absorption and emission maxima are each oppositely shifted from $\hbar\omega_{sr}$ by the photon recoil energy $\pm(\hbar k_{sr})/2M_0$, that the Stokes' shift between photon energies for the absorption and emission maxima is

$$(2\lambda) = \frac{\hbar^2 k_{sr}^2}{M_0}, \qquad (S4)$$

and that the absorption and emission lineshapes have the same photon energy variance

$$\Delta^2 = \frac{\hbar^2 k_{sr}^2 k_\mathrm{B} T}{M_0}. \qquad (S5)$$

Rewriting the variance in terms of the frequency gives the non-relativistic Doppler linewidth.[4] Substituting the lineshapes from Eq. (S3) into Eq. (7) of the main text, using the above change in standard chemical potential, rewriting the wavevector in terms of the frequency, and rearranging by completing the square shows that the absorption and emission lineshapes obey the generalized Einstein relation. The variance in Eq. (S5) and Stokes' shift between equilibrium absorption and emission in Eq. (S4) satisfy the generalized Einstein relation for homogeneous Gaussian lineshapes:[5] $\Delta^2 = (2\lambda)k_{\mathrm{B}}T$. The Stokes' shift required by the generalized Einstein relation equals the photon recoil shift. It should be noted that the simplified treatments of the Doppler lineshape presented here and in ref. 3 do not include the relativistic aberration and transformation of the electromagnetic energy density that go into the [first order in (*v*/*c*)] demonstration that the Einstein coefficients[6,7] and Einstein coefficient spectra[8] lead to an equilibrium Maxwell-Boltzmann molecular velocity distribution through molecule-radiation interaction alone.